\documentclass[a4paper,fleqn,usenatbib]{mnras}

\usepackage{newtxtext,newtxmath}
\usepackage[T1]{fontenc}
\usepackage{ae,aecompl}

\usepackage{graphicx}	% Including figure files
\usepackage{amsmath}	% Advanced maths commands

\usepackage{pdflscape}

\title[Turbulence analysis of M 42]{An analysis of the turbulence in the central region of the Orion Nebula (M42) - II. homogeneity and power-spectrum analyses}

\author[G. A. Anorve-Zeferino]{
G.~A. Anorve-Zeferino,$^{1,2}$\\
$^{1}$National Pedagogical University Unit 122 (UPN Unidad 122), Blvd. Vicente Guerrero S/N, 39715, Acapulco, Mexico\\ 
$^{2}$\'Ecole Polytechnique, Route de Saclay, 91120, Palaiseau, France\\
}

\date{Accepted XXX. Received YYY; in original form ZZZ}

\pubyear{2015}

\begin{document}
\label{firstpage}
\pagerange{\pageref{firstpage}--\pageref{lastpage}}
\maketitle

\begin{abstract}
In this second communication we continue our analysis of the turbulence in the Huygens Region of the Orion Nebula (M 42). We calculate the associated transverse structure functions up to order 8-th and find that the higher-order transverse structure functions are almost proportional to the second-order transverse structure function: we find that after proper normalisation, the higher-order transverse structure functions only differ by very small deviations from the second-order transverse structure function in a sub-interval of the inertial range. We demonstrate that this implies that the turbulence in the Huygens Region is quasi-log-homogeneous, or to a better degree of approximation, binomially weighted log-homogeneous in the statistical sense, this implies that there is some type of invariant statistical structure in the velocity field of the Huygens Region. We also obtain and analyse the power-spectrum of the turbulent field and find that it displays a large tail that follows very approximately two power-laws, one of the form $E(k)\propto k^{-2.7}$ for the initial side of the tail, and one of the form $E(k)\propto k^{-1}$ for the end of the tail. We find that the power-law with exponent $\beta\sim -2.7$ corresponds to spatial scales  of 0.0301--0.6450 pc. We find that the exponent of the first power-law $\beta\sim -2.7$ is related to the exponent $\alpha_2$ of the second-order structure function in the inertial range. We interpret the second power-law with exponent $\beta \sim -1$ as an indicator of viscous-dissipative processes occurring at scales of $\delta r=1$--5 pixels which correspond to spatial scales of 0.00043--0.00215 pc.
\end{abstract}


%%%%%%%%%%%%%%%%%%%%%%%%%%%%%%%%%%%%%%%%%%%%%%%%%%

%%%%%%%%%%%%%%%%% BODY OF PAPER %%%%%%%%%%%%%%%%%%

\section{Introduction}

\begin{figure}
	\centering
	\includegraphics[width=\columnwidth]{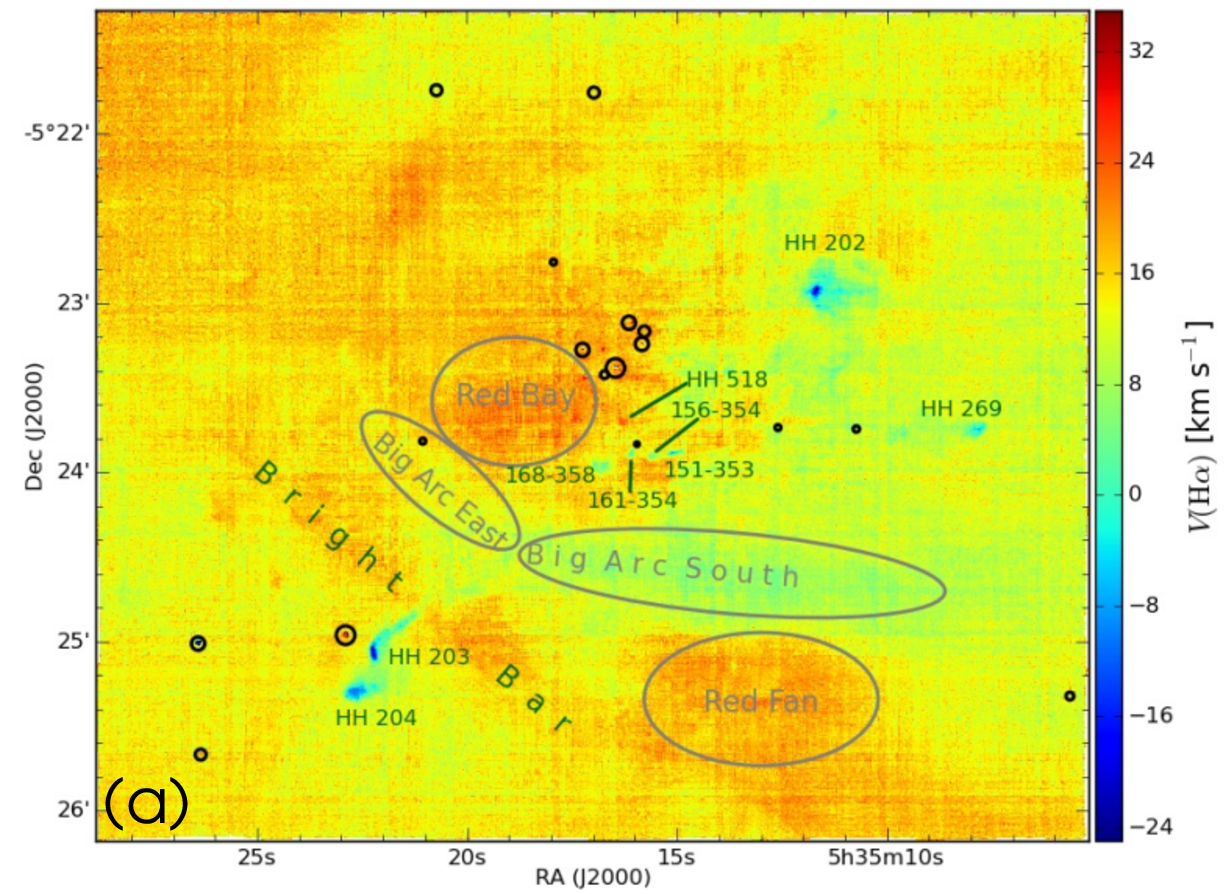}\\
	\includegraphics[width=0.94\columnwidth]{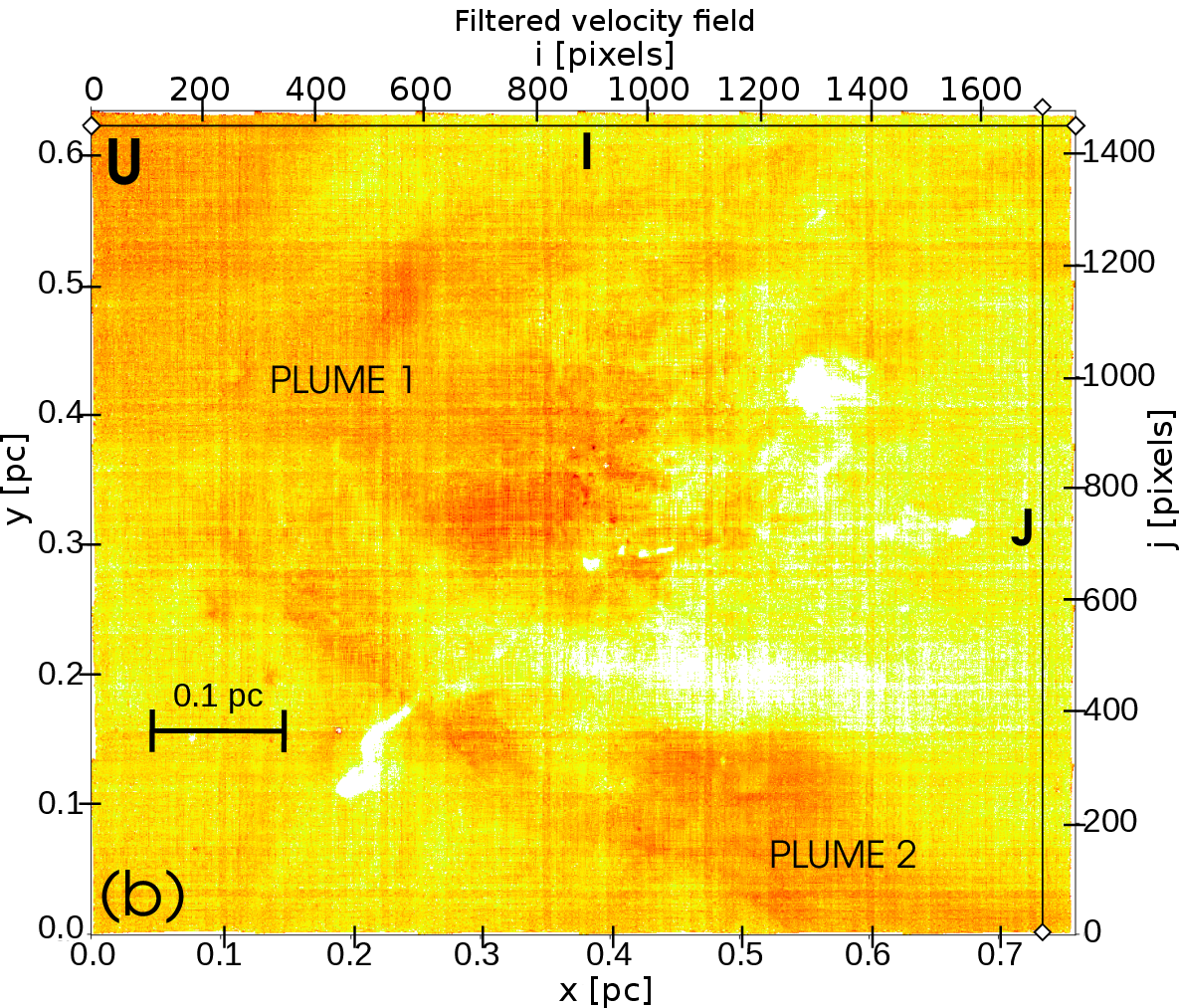}
	\caption{MUSE H$\alpha$ LOS centroid velocity field of Huygens Region of the Orion Nebula (M 42). Panel (a): figure 28(a) from \citet{Weilbacher2015} marking important projected flow structures called Red Bay, Red Fan, Bright Bar, Big Arc South and  Big Arc East. Panel (b): filtered image where we masked with white pixels the quiescent region which corresponds to the blue and green colored zones in Figure \ref{fig:fig1}(a). The masked zones have negative and small LOS centroid velocities below the -1$\sigma$ limit. Besides diffuse gas, the quiescent region contains also stars and Herbig-Haro objects. The extended turbulent field corresponds to the unmasked pixels and is labelled as $U$; it corresponds to the yellow and red colored pixels which represent intermediate and high LOS centroid velocities and to a small quantity of light green colored pixels surrounding the masked quiescent region, consult the colorbar in Figure~\ref{fig:fig1}(a). The axes ticks in Figure~\ref{fig:fig1}(b)  are given in parsecs and pixels.  I and J are the lateral sizes of the databox in pixels with I=1766 pixels and J=1476 pixels. Plume 1 and Plume 2 are projected turbulent regions that contain a significant percentage of the total power $k_2/2$ as these plumes are large and contain gas with high LOS centroid velocities.}
	\label{fig:fig1}
\end{figure}

\begin{figure}
	\centering
	\includegraphics[width=\columnwidth]{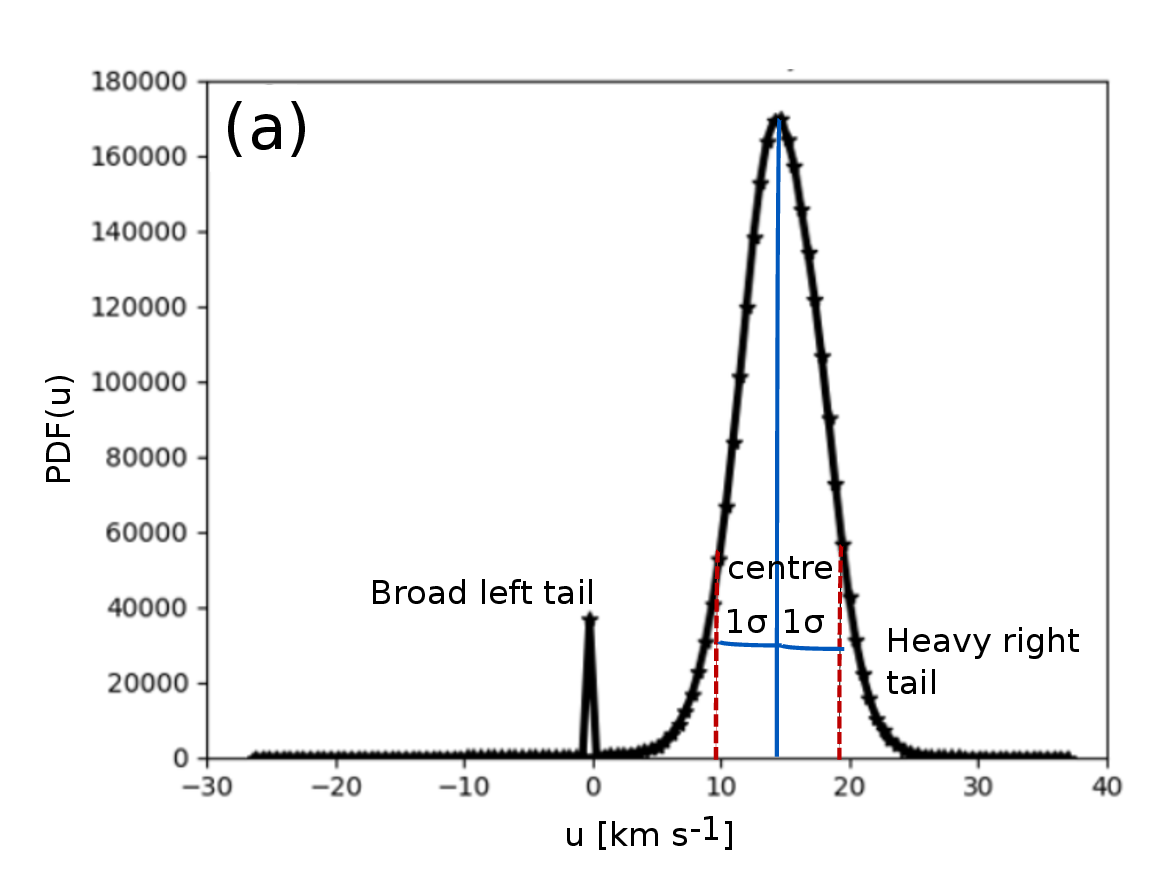}\\
	\includegraphics[width=\columnwidth]{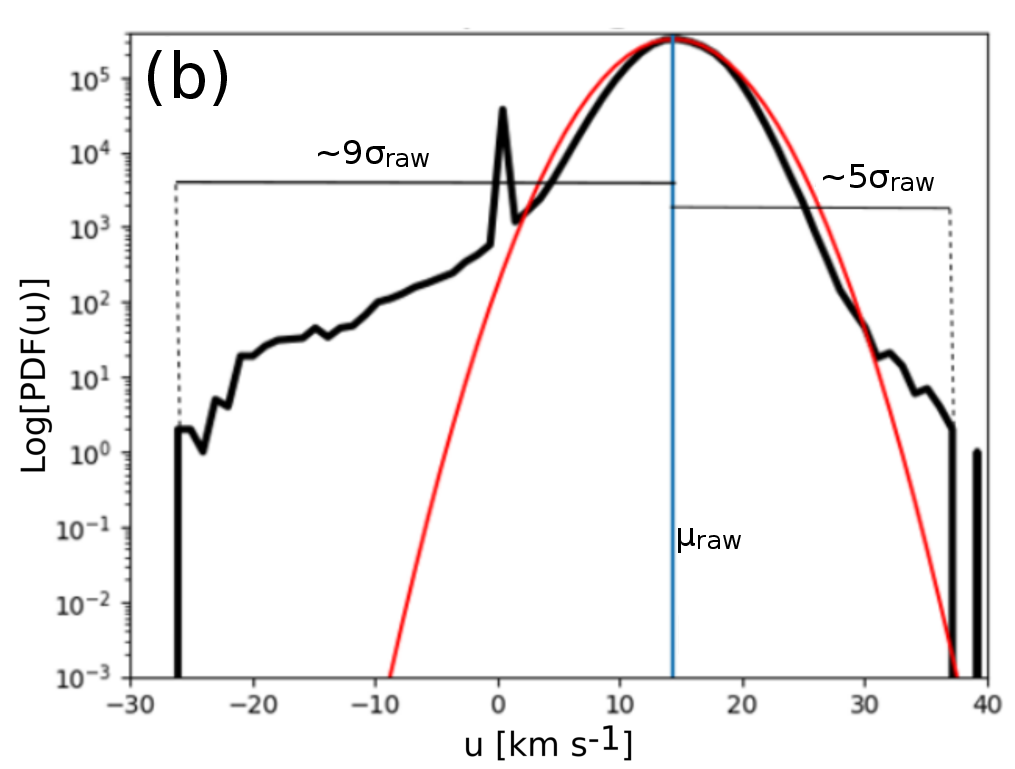}\\
	\caption{Non-normalised PDF of the Huygens Region between $-9\sigma$--$5\sigma$ limits, see Paper I. Panel (a): Non-normalised PDF in linear scale with different features labelled: the centre, the broad left-tail and the heavy right tail. We will identify which zones in the LOS centroid velocity map of the Huygens Region contribute to  each of these features. Panel (b): Non-normalised PDF in semi-logarithmic scale. Notice that the tails of the PDF are heavier that those corresponding to a Gaussian distribution. The red curve represents a Gaussian to which the broad centre of the non-normalised PDF was fitted and $\mu_{\rm raw}$ and $\sigma_{\rm raw}$ are the mean and the standard deviation of such Gaussian and also correspond to the mean and standard deviation of the LOS centroid velocity field without any filtering, see Paper I.}
	\label{fig:fig2}
\end{figure}

In a previous communication, \citet{AnorveZeferino2019}, henceforth Paper I, we analysed the turbulence in the Huygens Region of the Orion Nebula (M 42) through the non-normalised PDF and the transverse second-order structure functions corresponding to the \emph{MUSE} H$\alpha$ line-of-sight (LOS) centroid velocity map obtained by \citet{Weilbacher2015}. We found that the \emph{MUSE} H$\alpha$ line-of-sight (LOS) centroid velocity map of the Huygens Region of M 42 contains two components: a large extended turbulent region that contains 95.16\% of $k_2/2=\sum_{\rm i, j} u(\rm i,j)^2$,\footnote{$k_2$ is a particular case of $k_{\rm p}$ given by Eq.~(\ref{eqkp})} and a middle-sized elongated quiescent region that contains 2.14\% of $k_2/2$ and where young stars and Herbig-Haro objects reside, see Figure \ref{fig:fig1}. The term $k_2/2$ corresponds to the total power in the LOS centroid velocity map according to Parseval's theorem, $u$ is the LOS centroid velocity at a given pixel and (i,j) are the coordinates of that pixel in the LOS centroid velocity map. The rest of $k_2$/2, 2.70\% of it, is contained in very small sub-zones (a few pixels wide) around stars dispersed through the LOS centroid velocity map of the Huygens Region. We found that the transverse second-order structure function of the extended turbulent region has an inertial range that follows a power-law  of the form $S_2^{\rm w}\propto \delta r^\alpha$ with $\alpha=0.6824$ and $\delta r$ the projected separation distance. The previous exponent exceeds by less than two hundredths the Kolmogorov exponent for incompressible turbulence, $\alpha_{\rm K}=2/3\approx 0.6667$, which indicates that slightly compressible turbulence is ongoing in the Huygens Region. This conclusion is supported by the fact that the PDF of the LOS centroid velocity map of the Huygens Region is similar to the PDF of solenoidal turbulence obtained from high resolution numerical simulations of forced modal turbulence but with a broader left tail in logarithmic scale, compare Figure~\ref{fig:fig2}(b) with figure A1 in \cite{Federrath2013} and see also \citet{Stewart2022}. Solenoidal turbulence is the less violent form of forced modal turbulence which explains qualitatively the closeness of $\alpha$ to the Kolmogorov exponent $\alpha_{\rm K}$.

In Figure~\ref{fig:fig1}(a) we reproduce the \emph{MUSE} H$\alpha$ LOS centroid velocity map of the Huygens Region of M 42 obtained by \citet{Weilbacher2015} and below we present an image where the quiescent region was filtered out, Figure~\ref{fig:fig1}(b). The LOS centroid velocity field of the resulting extended turbulent region contais 95.16\% of $k_2/2$ and will be referred in what follows as $U$. We will analyse the turbulent field $U$, i.e. the extended turbulent region with the quiescent region removed by obtaining  and analysing the associated structure functions up to the order 8th and we will also obtain and analyse the associated power-spectrum. Structure functions have been calculated in the past in Astronomy and Astrophysics for different applications using Monte Carlo methods, e.g. \citet{Konstandin2012,Boneberg2015}, and spectral methods --related to the power spectrum-- including de-projection, e.g. \citet{Clerc2019,Cucchetti2019}. \emph{Pseudo-exhaustive}\footnote{Or better said, non-exaustive} real-space calculation methods have been of course also used to calculate structure functions, but they have yielded unsastifactory results, examples of this abound in the literature as remarked in Paper I where a revisitation of previous calculations of structure functions was concluded necessary. In this communication we use our new real-space weighted algorithm which includes necessary and sufficient zero-padding and exact circular averaging to calculate \emph{exhaustively}\footnote{i.e., using all the pixels in the LOS centroid velocity map} the transverse structure functions of the Huygens Region with the quiescent region removed and we use our new computational-geometry-based algorithm to calculate the associated power spectrum. The latter algorithm allows to  interpolate the power-spectrum satisfactorily and fit its long tail to the adequate functions, in this case two power-laws, Section~\ref{sec:sec4}.

For the sake of insight, we will first segment the LOS centroid velocity map of the Huygens Region in zones with differents hydrodynamics showing how their distinct dynamics reflect on the associated PDF. Our aim is to characterise graphically the turbulence on the Huygens Region which will produce a better understanding of the hydrodynamics; specifically, it will clarify which zones of the Huygens Region contribute the most to the turbulence and what features of the PDF of the LOS centroid velocity do they produce, see Figure~\ref{fig:fig2}. The results of this analysis are discussed below and summarised on Table~\ref{tab:tab1}.

First, in the filtered image that we will analyse, Figure~\ref{fig:fig1}(b), the Big Arc South  was filtered out together with the gas that surrounds a set of young starts and Herbig-Haro objects to the North-West as well as a proximate nebulosity to the South; a 'bullet' containing the jets of the Herbig-Haro objects HH 203 and HH 204 was also filtered out, compare with Figure~\ref{fig:fig1}(a). Overall, these areas constitute the extended quiescent region that we filtered by considering turbulent and coupled only the data that corresponds to the PDF of the LOS centroid velocity of the Huygens Region of the Orion Nebula within -1$\sigma$--5$\sigma$ limits, see Paper I. Evidently, the extended turbulent region $U$ [non-masked pixels in Figure~\ref{fig:fig1}(b)] contains the highest velocities in the map as well as the flow patterns of interest. In counterpart, the gas in the quiescent region contains smaller and negative projected velocities with a mean of $\mu_u=6.65$ km s$^{-1}$ --which is 56\%  smaller than that of the extended turbulent region $U$-- and a velocity dispersion of $\sigma_u=3.70$ km s$^{-1}$. \emph{The quiescent region produces the broad left tail of the LOS centroid velocity PDF  of the Huygens Region of the Orion Nebula}, consult Figure 2 and figures 3(a) and 3(b) on Paper I. However, in Paper I, we determined that the quiescent region has an uncoupled dynamics from the rest of the velocity field; this manifests through the presence of a secondary peak in the PDF just to the left of the mean, see Figure~\ref{fig:fig2}, which indicates a distribution with different properties affecting and producing most of the left tail of the main PDF. The uncoupled dynamics of the quiescent regions is probably related to the presence of young stars, proto-stars and Herbig-Haro objects that "encapsulate" the three-dimensional region corresponding to the quiescent region through the effect of winds,\footnote{See the polytropic free wind model in \cite{AnorveZeferino2009}} jets and radiation which suppress the turbulence in the quiescent region establishing a quasi-uniform quiescent velocity field, see figure 10 in \citet{Weilbacher2015} where this can be clearly seen in several line fluxes and see also figure 5(b) and table 1 in Paper I where we showed that the exponent of the power-law to which the second-order structure function of the quiescent region alone can be fitted is very small, $\alpha_{\rm q}=0.0853$, which indicates an homogeneous non-turbulent velocity field statistically.

\begin{figure}
	\centering
	\includegraphics[width=0.95\columnwidth]{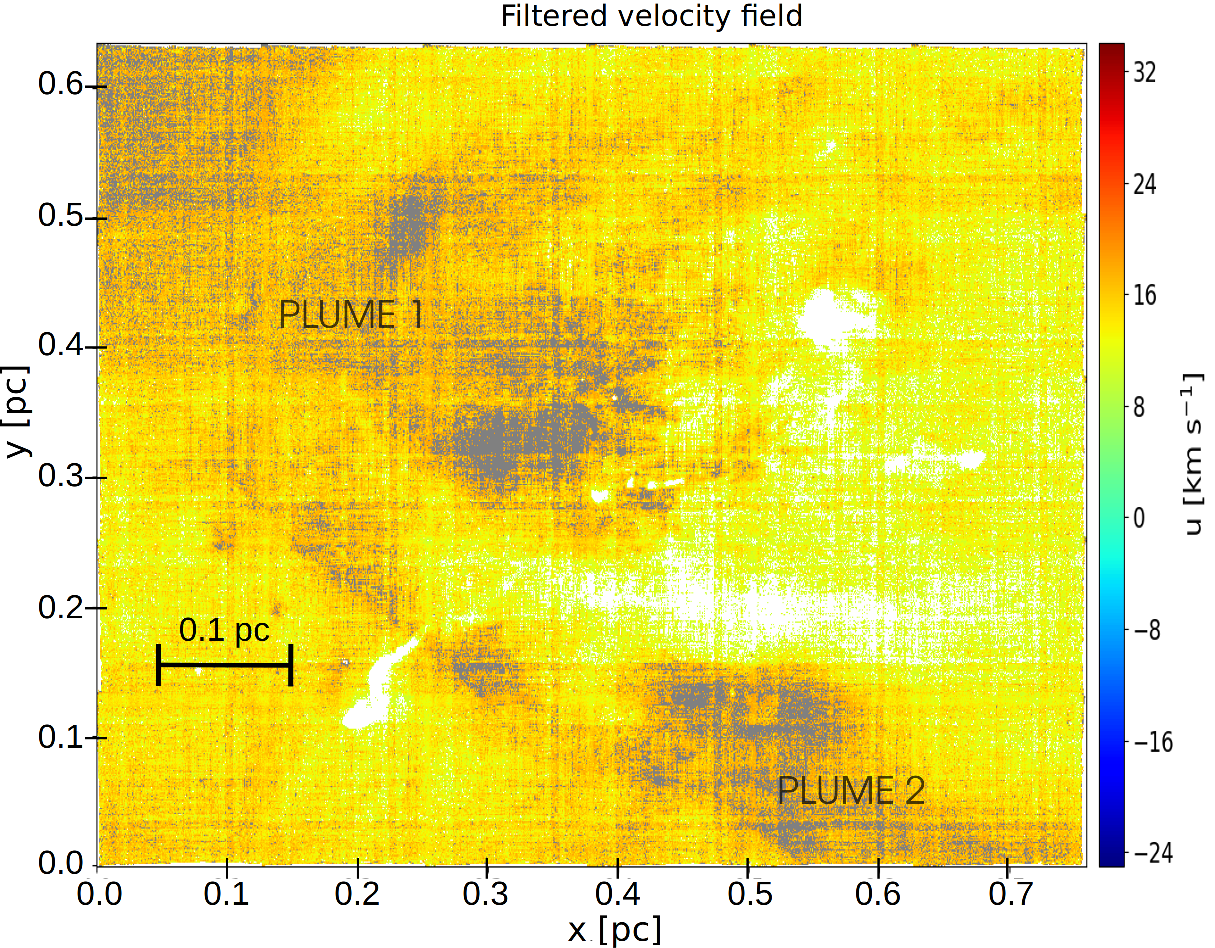}\\
	\caption{MUSE H$\alpha$ LOS centroid velocity map of the Huygens Region of the Orion Nebula with the data that exceeds -1$\sigma$--1$\sigma$ limits of the PDF of the whole region masked. The quiescent region is masked with white pixels and the data within 1$\sigma$--5$\sigma$ limits is masked with grey pixels. A significant fraction of Plume 1 is now masked with gray pixels as well as a large fraction of Plume 2; additionally, most of the Red Bay and the Red Fan are now masked with gray pixels as well as a significant part of the Bright Bar [compare with Figures~\ref{fig:fig1}(a)-(b)];  these zones which have been masked with gray pixels produce the heavy right tail of the LOS centroid velocity PDF of the Huygens Region of the Orion Nebula --which is an indicator of ongoing turbulence-- and contribute with $\approx 16\%$ of $k_2/2$ (Paper I). The  non-masked regions are also turbulent and contain approximately 79\% of $k_2/2$. These non-masked regions produce the central interval of the PDF without any tails, see Figure~\ref{fig:fig2}.}
	\label{fig:fig3}
\end{figure}

Hence, the turbulence in the Huygens Region of the Orion Nebula is comprised in most of the image except by the elongated quiescent region masked in  Figure \ref{fig:fig1}(b) and a few regions a few pixels wide where sometimes stars and proto-stars were detected to reside. On the other hand, the LOS centroid velocity field $U$ of the extended turbulent region [non-masked pixels in Figure \ref{fig:fig1}(b)] has a mean projected velocity of $\mu_u=15.11$ km s$^{-1}$ and a velocity dispersion of $\sigma_{u}=2.80$ km s$^{-1}$. This turbulent field includes three strongly marked features: the Bright Bar, the Red Fan and the Red Bay, see Figure \ref{fig:fig1}(a); additionally, there are two large high-velocity projected plumes crossing the field, one to the North-East of the Red Bay and one to the South-West of the Red Fan which have been labelled respectively as Plume 1 and Plume 2 in Figure \ref{fig:fig1}(b). \emph{These three strongly visible structures --The Bright Bar, Red Fan and Red Bay-- together with a significant part of the the plumes --but not the full plumes-- produce the right heavy tail of the LOS centroid velocity PDF of the Huygens Region of the Orion Nebula} [see also Paper I, section 2-Filter 4 and figures 3(b) and 3(d)] and they contribute with $\sim 16$\% of the total power, $k_2/2$, of the field; see Figure~\ref{fig:fig3} which present these regions masked with gray pixels. The turbulence is stronger in these regions since they produce the heavy right tail of the PDF of the LOS centroid velocity field of the Huygens Region, tail which corresponds to the highest velocities. The presence of the  right heavy tail in the PDF of the LOS centroid velocity is an indicator that turbulence is ongoing in the field $U$, see \citet{Federrath2010,Federrath2013} and \citet{Stewart2022}.  

Approximately 79\% of the total power in the LOS centroid velocity map, $k_2/2$, is contained on a large area of the  North-East plume (Plume 1), an intermediate percentage area of the South-West plume (Plume 2) and in the extended yellow zones on the map of the Huygens Region, see Figure \ref{fig:fig3}. \emph{This unmasked zones in Figure~\ref{fig:fig3} produce the central body (the central interval within -1$\sigma$--1$\sigma$ limits which excludes the tails) of the PDF of the LOS centroid velocity of the Huygens Region}.

Thus, through the analysis of the filtered field $U$ shown in Figure~\ref{fig:fig1}(b) we will analyse the zone that contains most of the total power in the LOS centroid velocity map as well as the turbulent features. A resume of the zones that compose the Huygens Region giving the percentage of $k_2/2$ that they contain as well as the features of the PDF that they produce is given in Table~\ref{tab:tab1}.

\begin{table}
	\centering
	\caption{Identified zones in the Huygens Region of the Orion Nebula using our PDF filtering. First column: zone name; second column: percent of the total power, $k_2/2$, on the zone; third column: feature of the PDF of the LOS centroid velocity map of the Huygens Region which the zone produces.}\label{tab:tab1}
	\begin{tabular}{lcc} % four columns, alignment for each
		\hline
		Zone(s)                      & \% of $k_2/2$    & Feature of the PDF     \\
		                             &                &                        \\
		\hline
		1. Turbulent region $U$      & 95.16\%        & centre+right heavy tail\\                                 
		2. Quiescent region          & 2.14\%         & broad left tail    \\
		3. Very high velocity zones  & 2.70\%         & filtered from the PDF  \\
		with a width of a few pixels &                &\\
		4. Most of the Red Fan,      &                &\\
		Red Bay and Bright Bar and   &                & \\                       
		part of Plume 1 and Plume 2  & 16.48\%        & right heavy tail       \\
		5. Unmasked zones in         &                &                        \\
		Figure~\ref{fig:fig3}        & 78.68\%        & centre without tails   \\
		\hline
	\end{tabular}
\end{table}

The rest of the Paper is organised as follows. In Section \ref{sec:sec2} we will first give the theoretical framework on which our calculations of the transverse structure functions and the power-spectrum are based. This Section includes definitions as well as normalisation and convergence results that are important to understand better the concept and properties of structure functions. Section \ref{sec:sec2} should be of interest because our methods improve previous algorithms. The reader only interested on the outcomes of our calculations and the astrophysical implications derived from our analyses may omit this Section and pass directly to Section~\ref{sec:sec3}.  In Section~\ref{sec:sec3} we obtain the transverse structure functions up to the 8-th order that correspond to the extended turbulent region $U$ and find that the higher-order structure functions are almost proportional to $S_{2}^{\rm w}(\delta r)$, i,e. that they differ from $S_{2}^{w}$ almost only by a constant multiplicative factor. In Section \ref{sec:sec4} we obtain the power-spectrum of the extended turbulent region. We find that the power-spectrum has a long tail  which can be fitted to two power-laws.  Both power-laws are robust since they are an outcome of our exact computational geometry algorithm. In Section~\ref{sec:sec5} we interpret the physical meaning of the relation between the higher-order transverse structure functions and also analyse the power-spectrum. Our conclusions are given in Section~\ref{sec:sec6}.

\section{Theoretical framework}\label{sec:sec2}

\begin{figure}
	\centering
	\includegraphics[width=0.675\columnwidth]{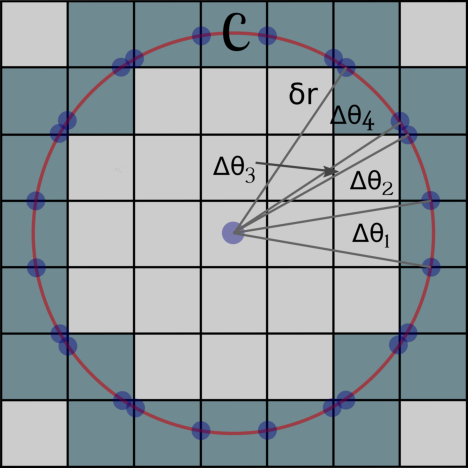}\\
	\caption{Circumference $\mathcal{C}$ of radius $\delta r=3$ pixels traced surrounding an arbitrary pixel whose centre has been marked with a wide blue dot. The circumference crosses exactly 8$\delta r=24$ pixels highlighted in gray-blue. Four distinct subtended angles $\Delta\theta_{\rm i}$ (for this value of $\delta r$) are shown, they repeat in a periodic cyclic pattern until 8$\delta r$ subtended angles are completed.}
	\label{fig:fig4}
\end{figure}

We assume that the line of sight emission from the Huygens Region is completely perpendicular to the observation plane, see figure 5 in \citet{AnorveZeferino2009}. Under this planar approximation, the p-th order weighted transverse structure function can be simply defined for \emph{even} p as

\begin{equation}
	S_{\rm p}^{\rm w}(\delta r) = \left< [u({\bf r}+\delta{\bf r}) - u({\bf r})]^{\rm p}\right>_{\rm w}\label{genw}
\end{equation}

\noindent  and for \emph{odd} p as

\begin{equation}
	S_{\rm p}^{\rm w}(\delta r) = \left< |u({\bf r}+\delta{\bf r}) - u({\bf r})|^{\rm p}\right>_{\rm w}\label{genw}
\end{equation}

\noindent where p is the order of the structure function, the superscript w indicates that the structure function is calculated using weights, u is the LOS centroid velocity, $\bf{r}$ is the position vector that indicates the location of the pixel with velocity u in the computational grid, $\delta{\bf r}$ is the displacement vector in pixels, $|\,|$ indicates the absolute value\footnote{We do not take the absolute value of the difference of velocities in the case of the even-order structure functions because is net effect is null because of the parity of p and because by omitting taking the absolute value the algorithm is faster for even p}  and the angle brackets $<\,>_{\rm w}$ indicate a pixel-by-pixel weighted average over $\mathbb{R}^2$, see also \citet{Babiano1985} and \citet{Thomson1988}. The weighted average is carried out as follows: using our computational geometry algorithm with trace an exact circumference $\mathcal{C}$ of radius $\delta r$ around each pixel and define the weights for the pixels on $\mathcal{C}$ as

\begin{equation}
	w_{\rm i}=\frac{\Delta\theta_{\rm i}}{2\pi}
\end{equation}

\noindent where $\Delta\theta_{\rm i}$ is the angle subtended by each pixel intersected by $\mathcal{C}$ and  i is a labelling index that identifies those pixels, see Figure \ref{fig:fig4} which shows a scheme of this geometrical construction. The weights ponder the contribution of each pixel to the structure function in an exact manner and allow to take advantage of the full resolution of the grid. Because of this our calculations are exact up to machine numerical error. We proved analytically, geometrically/graphically and also verified numerically that a circumference $\mathcal{C}$ of radius $\delta r$ intersects exactly 8$\delta r$ pixels and we use this fact to optimise our algorithms.

The bidimensional p-th order weighted transverse structure function can then be calculated \emph{exhaustively} through the formulae

\begin{equation}
S_{\rm p}^{\rm w}({\delta {r}}) =  \sum_{\text{m,n}}^{\text{M'',N''}} \sum_{\theta} {\rm w}(\theta;\delta r)\{u[{\bf r}+{\delta \bf{r}}(\theta)]- u({\bf r})\}^{\rm p}\label{eqweighs_gcoords}
\end{equation}

\noindent for even p and 

\begin{equation}
	S_{\rm p}^{\rm w}({\delta {r}}) =  \sum_{\text{m,n}}^{\text{M'',N''}} \sum_{\theta} {\rm w}(\theta;\delta r)|u[{\bf r}+{\delta \bf{r}}(\theta)]- u({\bf r})|^{\rm p}\label{eqweighs_gcoords_odd}
\end{equation}

\noindent for odd p, where $\theta$ is the polar angle and the angular sum is a discrete sum that needs to be carried taking into account the lower and upper limits of the angles subtended by  each pixel crossed by $\mathcal{C}$ since their difference provides the magnitude of $\Delta\theta_{\rm i}$. Such limits have been indicated by wide blue dots over $\mathcal{C}$ in Figure \ref{fig:fig4}. Notice that the weights are a function of both the polar angle and the displacement radius and thus they are different for distinct values of $\delta r$. 

We will define the indices m, n, M'' and N'' of the outer summation in Eq. (\ref{eqweighs_gcoords}) and Eq.~(\ref{eqweighs_gcoords_odd}) in the next Subsection. These outer summations indicate summation over a grid extended through zero-padding over which the transverse structure function will be calculated and because of this it is very important to define properly the limits m, n, M'' and N'' since they contribute not only the exactness of the structure functions but  also determine the speed of the algorithm. An optimal choosing of these limits can avoid unnecesary large calculation times.

\subsection{Integration limits for calculating $S_{\rm p}^{\rm w}(\delta r)$, optimal zero-padding}\label{sec:sec2.1} 

In the continuous case, just as the circular weighted correlation function $C^{\rm w}(\delta r)$ that also depends on the separation radius $\delta r$ and to which the second-order structure function is closely linked [see \citet{Schulz-Dubois1981}], the  weighted bidimensional p-th order transverse structure function \emph{of a field $U$} is also \emph{a function defined through integration over all space} when calculated through circular averages in real space, its definition in the continuous case is:

\begin{equation}
S_{\rm p}^{\rm w}(\delta r)=\frac{1}{2\pi }\int_{-\infty}^{\infty}\int_{-\infty}^{\infty}\int_{0}^{2\pi} |u[({\rm x,y})+{\delta \bf{r}}(\theta)]- u({\rm x,y})|^{\rm p} {\rm d}\theta  {\rm d}x {\rm d}y.\label{Spcont2D}
\end{equation}

\noindent  For  p=2, the integration over all space makes the result of the calculation in real space through circular averages \emph{to coincide exactly}\footnote{See figure 5 in \citet{AnorveZeferino2019}} with the result obtained from  the Fourier-transform-based algorithm; see v.gr. \citet{Batchelor1949,Batchelor1951,Monin1971,Schumacher1994,ZuHone2016,Cucchetti2019} and \citet{Clerc2019} for the definition and calculation details of the second-order structure function $S_{2}(\delta r)$ through Fourier-transform-based methods. The Fourier-transform-based theory yields the relationship

\begin{equation}
S_{2}^{\rm w}(\delta r)=2[C^{\rm w}(0)-C^{\rm w}(\delta r)],\label{eqS2w}
\end{equation}

\noindent  which is a classical result that is well-known to always yield the correct results. In the previous equation $C^{\rm w}(\delta r)$ is the weighted circular correlation function. Notice that in \citet{Batchelor1949,Batchelor1951,Monin1971,Schulz-Dubois1981,Schumacher1994,ZuHone2016,Cucchetti2019} and \citet{Clerc2019}, the correlation function and thus the second-order structure functions are not weighted as in this communication, i.e. as indicated in Eq.~(\ref{eqS2w}), however, we have proved analytically that the relation holds also in the weighted case here presented.

Hence, in order to calculate $S_{\rm p}^{\rm w}(\delta r)$ in the discrete case for uniform Cartesian grids through Eqs.~(\ref{eqweighs_gcoords})~and~(\ref{eqweighs_gcoords_odd})\footnote{Notice that Eqs.~(\ref{eqweighs_gcoords})~and~(\ref{eqweighs_gcoords_odd}) have been normalised by the pixel area $a={\rm d}x{\rm d}y=({\rm d}x)^2=({\rm d}y)^2$ to suppress its contributrion in order to make the definitions more general} it is compulsory to zero-pad the databox that stores the field $U$ to account properly for integration over all space. Given that the field $U$, as astrophysical velocity fields in general, has finite extension, then, for a given $\delta r$ the integration over all space $\mathbb{R}^2$ reduces in practice to integration over a grid extended with finite zero-padding. In order to illustrate this, consider for instance the calculation of the 1D correlation function of two identical rectangle functions of width $b$ pixels. In this case, although the definition of the correlation function implies integration over all space, the correlation does not vanish only in a finite interval of width equal to 3$b$ pixels; thus only finite zero-padding\footnote{A zero-padding of $b$ pixels to each side of the fixed rectangle function of width $b$ pixels} is needed to calculate the correlation function where it does not vanish. An analogue situation occurs for structure functions when calculated through circular averages in real space, only a very specific amount of zero-padding is required to calculate them. Using this specific optimal amount of zero-padding for each $\delta r$ can save substantial calculation time without affecting the exactness of the calculation.

Thus, as in the case of the correlation function, the calculation of $S_{\rm p}^{\rm w}(\delta r)$ has to be done over a domain that consists of the original databox containing the field $U$ plus necessary and sufficient zero-padding for each $\delta r$.\footnote{Notice that in order to calculate $S_{\rm p}^{\rm w}(\delta r)$ for a given field $U$ \emph{uncoupled} nearby and far away fields appearing in the velocity map need to be masked with zeroes.} When less than this necessary and sufficient zero-padding is added, edge effects  degrade the structure functions \citep{AnorveZeferino2019}; e.gr. for p=2, when less than the necessary and sufficient zero-padding is added to calculate through Eq.~(\ref{eqweighs_gcoords}) the results of the calculation will not coincide with the exact results obtained through the Fourier-transform-based algorithm summarized in Eq.~(\ref{eqS2w}) because of edge effects. When more than the necessary and sufficient zero-padding is added the calculation is inefficient because more circular averages than necessary will be carried out. We give analytical expressions for the optimal amounts of zero-padding to calculate $S_{\rm p}^{\rm w}(\delta r)$ below, these amounts are obtained from purely and simple geometrical considerations that ensure that all the required circular averages to account for integration over all space are carried out. We omit here the statement of such considerations and their related derivations because of space reasons, we only present the final result.

So, let $\delta r^{**}$ be \emph{the maximum separation distance for which the structure function $S_{\rm p}^{\rm w}(\delta r)$ will be calculated} and I$\times$J the area in pixels$^2$ of the unpadded databox that contains the field $U$, for the case of the map of the Huygens Region analysed in this communication I$\times$J=1766$\times$1476 pixels$^2$. If one assumes that the unpadded databox is completely filled with data,\footnote{If this is not the case one should first reduce the unpadded databox to the most compact databox that contains the non-vanishing data to analyse} then, for maximum efficiency and analytical sufficiency (i.e. to ensure correct and exact results), when one is to calculate the mapping $S_{\rm p}^{\rm w}(\delta r)$ up to a maximum displacement radius $\delta r^{**}$, zero-padding must be added \emph{symmetrically} up to obtaining an area equal to $\mathcal{A}_{\rm ZP}(\delta r^{**};{\rm I,J})$ given by

\begin{equation}
\mathcal{A}_{\rm ZP}(\delta r^{**};{\rm I,J})=  ({\rm I}+4\delta r^{**})\times({\rm J}+4\delta r^{**})\,\mbox{pixels}^2={\rm M'}\times {\rm N'}\,\mbox{pixels}^2.\label{eqZP1}
\end{equation}

\noindent With 'symmetrically' we mean that each side of the I$\times$J databox needs to be extended adding zero-padding areas with a width of $2\delta r^{**}$ pixels zero-padding also the areas associated to the four corners to obtain an extended rectangular databox of area $\mathcal{A}_{\rm ZP}(\delta r^{**};{\rm I,J})$. In practice, this can be done easily by creating a 2D array of size $A_{\rm ZP}(\delta r^{**};{\rm I,J})$ and centring the unpadded databox that contains the data to analyse in such array, see Subsection~\ref{sec:sec2.4}.

After zero-padding as indicated by Eq~(\ref{eqZP1}), for carrying optimally the respective calculation of $S_{\rm p}^{\rm w}(\delta r)$ \emph{for smaller or equal} radii $\delta r'\le\delta r^{**}$, the area of the previous extended grid must be reduced \emph{logically} and \emph{symmetrically} for every $\delta r'$ to $\mathcal{A}_{\rm Int}(\delta r',{\rm I,J})$, which is given by

\begin{equation}
\mathcal{A}_{\rm Int}(\delta r';{\rm I,J})=({\rm I}+2\delta r')\times({\rm J}+2\delta r')\,\mbox{pixels}^2. \label{eqZP2}
\end{equation}

\noindent In this case 'symmetrically' means that the databox extended through Eq.(\ref{eqZP1}) must be reduced \emph{logically} to a sub-databox that contains the original I$\times$J databox extended considering zero-padding areas with a width of $\delta r'$ pixels to each of the unpadded databox sides zero-padding also the areas associated to the four corners to obtain a rectangle of area $\mathcal{A}_{\rm Int}(\delta r';{\rm I,J})$. 'Logically' means that this must be done by using array indices, i.e. without reducing physically the size of the array that contains the databox extended to the area  $\mathcal{A}_{\rm ZP}(\delta r^{**};I,J)$, Eq.~(\ref{eqZP1}).\footnote{Because part of the pixels in $\mathcal{A}_{\rm ZP}(\delta r^{**};{\rm I,J})$ and outside the area $\mathcal{A}_{\rm Int}(\delta r';{\rm I,J})$ will be used as support for circles of radius $\delta r'$} Such array indices are given below in Eq.~(\ref{eqMN}) and they correspond to the lower limits m,n and the upper limits M'' and N'' of the outer sommatorias in Eq.  (\ref{eqweighs_gcoords}) and Eq.~(\ref{eqweighs_gcoords_odd}) \emph{for each $\delta r'\le \delta r^{**}$}. These indices are:

\begin{equation}
\left.
\begin{array}{ll}
\text{m}&=2\delta r^{**}-\delta r',\\
\text{n}&=2\delta r^{**}-\delta r',\\
\text{M''}&=2\delta r^{**}+(\text{I}-1)+\delta r', \\[2pt]
\text{N''}&=2\delta r^{**}+(\text{J}-1)+\delta r';     
\end{array} \right. \label{eqMN}
\end{equation}

\noindent where we have assumed that the two-dimensional array that contains the extended databox of area $\mathcal{A}_{\rm ZP}(\delta r^{**};I,J)$ begins at the pixel coordinate $(i,j)=(0,0)$ which for simplicity we have assumed to correspond to either the lower or upper left corner pixel of the extended grid. \emph{We will assume that this is always the case in what follows}. With the indices indicated in Eq.~(\ref{eqMN}) one can calculate $S_{\rm p}^{\rm w}(\delta r)$ for all $\delta r=\delta r'\le\delta r^{**}$ using Eq.~(\ref{eqweighs_gcoords}) or Eq.~(\ref{eqweighs_gcoords_odd}). This method of calculation ensures that the calculation is \emph{exhaustive}, \emph{exact} and \emph{optimal regarding computational time} which is important since many previous works in the literature have used inadequate and imprecise methods of calculation and reported incorrect structure functions; as we remarked in Paper I, a revisitation of previous calculations may be necessary to obtain adequate diagnoses of previously analysed turbulent fields.

\subsection{Absolute/Universal normalisation of the structure functions}\label{sec:sec2.2}

In general, Equations~(\ref{eqweighs_gcoords}) and (\ref{eqweighs_gcoords_odd}), which allow to calculate exhaustively and exactly the weighted structure functions,\footnote{The Equations allows to calculate the non-weighted structure functions also by setting all weights to one  (i.e. by setting $w_{\rm i}=1$ for all i) and normalising instead by $8\delta r$; however, the weighted algorithm is exact for all $\delta r$ unlike the non-weighted version} will yield different asymptotic limits for different velocity fields $U$ and/or exponents p, i.e. different limits for large enough displacement radii $\delta r$. However, here we will introduce a \emph{normalisation factor} for each p-th order structure function that is absolute in the sense of convergence.\footnote{Absolute in the sense of convergence roughly means that is exact for $\delta r$ larger than a very specific value $D$ and that the normalisation factor normalises properly the structure functions for all $\delta r$} This universal normalisation factor will allow to compare in a convenient manner a) structure functions of different orders $p$ for the same velocity field $U$ and b) structure functions of the same order p for different velocity fields $U$. \emph{The former can be useful to find scaling laws for structure functions of the same field $U$ and the latter to test the often proposed universality of turbulence,\footnote{If turbulence were universal, structure functions of the same order p for different turbulent velocity fields would coincide almost everywhere when normalised properly} for instance}. In a work in preparation, hereafter Paper II, we have found analytically that there is an absolute/universal normalisation factor $k_{\rm p}$ for each order p which expression is indeed very simple: 

\begin{equation}
	k_{\rm p}=2\sum_{\rm q \in \mathcal{L}_0} |u_{\rm q}|^{\rm p}=2\sum_{\rm (i,j) \in \mathcal{L}_0} |u(i,j)|^{\rm p};\label{eqkp}
\end{equation}

\noindent where $\mathcal{L}_0$ is the finite spatial domain where the velocity field $U$ is defined, i.e. the unpadded databox of size I$\times$J, $|\,|$ indicates the absolute value,\footnote{Notice that taking the absolute value is only relevant when p is odd and there are negative velocities in the velociy field} $u_{\rm q}$ is the velocity at a given pixel, and q indicates the pair of coordinates $(i,j)$ on $\mathcal{L}_0$ of that pixel. In the case of the map  of the Huygens Region of the Orion Nebula that we will analyse the domain $\mathcal{L}_0$ is an uniform Cartesian grid of area I$\times$J=1766$\times$1476 pixels$^2$. Notice that for $p=2$ the normalisation factor $k_{\rm p}$ corresponds to two times the total power on the LOS centroid velocity map.

Thus, we define the \emph{normalised} pth-order weighted transverse structure function as 

\begin{equation}
S_{\rm p,n}^{\rm w}({\delta {r}}) = \frac{1}{k_{\rm p}} \sum_{\text{m,n}}^{\text{M'',N''}} \sum_{\theta} {\rm w}(\theta;\delta r)\{u[{\bf r}+{\delta \bf{r}}(\theta)]- u({\bf r})\}^{\rm p}\label{eqweighs_norm}
\end{equation}

\noindent for even p and

\begin{equation}
	S_{\rm p,n}^{\rm w}({\delta {r}}) = \frac{1}{k_{\rm p}} \sum_{\text{m,n}}^{\text{M'',N''}} \sum_{\theta} {\rm w}(\theta;\delta r)|u[{\bf r}+{\delta \bf{r}}(\theta)]- u({\bf r})|^{\rm p};\label{eqweighs_norm_odd}
\end{equation}

\noindent for odd p, where the subscript n indicates normalisation and the only difference with Eqs.~(\ref{eqweighs_gcoords}) and (\ref{eqweighs_gcoords_odd}) is the normalisation factor $k_{\rm p}$.

\emph{This way normalised all p-th order structure functions have the next nice property:}

\begin{equation}
S_{\rm p,n}^{\rm w}(\delta r)\rightarrow 1,
\end{equation}

\noindent for large enough separation distances $\delta r$ independently of the nature of the field $U$ in the unpadded databox.  We will prove numerically the above mentioned property of the transverse structure functions in Section \ref{sec:sec3}.

The absolute/universal convergence of $S_{\rm p}^{\rm w}(\delta r)$ to 1  after normalisation by $k_{\rm p}$ can be described more precisely: we will put forward below a simplified version of a theorem on Paper II which gives the exact separation distance $\delta r$ at which the normalised structure functions converge absolutely to 1, namely $\delta r=\delta r^*=D$. Let $D$ be the maximum linear separation distance present in the field $U$ defined for pairs $u(x_{\rm j},y_{\rm j})$ and  $u(x_{\rm i},y_{\rm i})$  of non-vanishing velocities in the unpadded databox $\mathcal{L}_0$ through the Eulerian/Cartesian metric, i.e.:

\begin{equation}
D=\left\lceil\max\left(\left\{[(x_{\rm j}-x_{\rm i})^2+(y_{\rm j}-y_{\rm i})^2]^{1/2}\right\}_{(\rm i,j)}\right)\right\rceil\label{eqEM}
\end{equation}

\noindent \emph{where $x$ and $y$ are the coordinates $(x,y)$ in units of pixels of each pair of non-vanishing velocities on $U$, $\lceil\,\rceil$ indicates the ceil operator\footnote{The ceil operator compensates exactly for the discrete nature of the uniform Cartesian grids and it yields an exact result in terms of whole pixels units} and max indicates the maximum operator which simply selects the maximum value in a set $\{\,\}$. The subindexes i and j must cover all pixels  that contain non-vanishing velocities such that one could obtain the maximum linear separation distance $D$ after applying the maximum operator to all calculated Eulerian/Cartesian metrics.}\footnote{Notice that the linear distance corresponding  to $D$ always connects two pixels adjacent to the contour of the domain that contains $U$. Here, for simplicity, we avoided  a definition of $D$ in term of contour points, however, notice that it can be much more convenient computationally, specially for large domains/grids} Notice that when the unpadded databox is completely filled with data $D=L$, where $L$ is the diagonal of the unpadded databox, i.e. $L=\lceil\sqrt{I^2+J^2}\rceil$. The referred theorem in Paper II states that

\begin{equation}
S_{\rm p,n}^{\rm w}(\delta r\ge D)=1,
\end{equation}

\noindent which implies that all normalised structure functions are \emph{perfectly flat} for all $\delta r\ge D$, i.e. they exhibit a perfectly flat plateau for $\delta r\ge D$. This occurs because all the correlations vanish for $\delta r\ge D$, see for instance Eq.~(\ref{eqS2w}) for the case p=2 and also Section~\ref{sec:sec3} and Section~\ref{sec:sec5.1}. All the plateaus are at the level of 1 because of the normalisation by $k_{\rm p}$.\footnote{Obviously, without normalisation the plateaus are at the level of the different $k_{\rm p}$'s} These flat plateaus at the level of 1 (or $k_{\rm p}$ if the transverse structure functions are not normalised) can be used to verify the correctness of the calculation of the structure functions. Also, with the structure functions properly normalised by $k_{\rm p}$, we will be able to compare structure functions of different orders p for the same field $U$ more adequately to find scaling laws and we will be also able to compare structure functions of the same order p for different velocity fields to test universality or other properties of turbulence. This will occur because the numerical values of the structure functions, and thus of their slopes included those of the inertial ranges, will be now scaled adequately, i.e. they will now have more similar and thus more easily comparable values. We will prove how advantageous is the normalization by $k_{\rm p}$ in Section~\ref{sec:sec3}, where we will calculate the normalised transverse structure functions up to the 8-th order. The advantages will become more patent in Section~\ref{sec:sec5} were we will interpret our results and prove that they imply that the turbulent field $U$ of the Huygens Region has a special type of homogeneity.

\subsection{{Power spectrum}}\label{sec:sec2.3}

In practice, the radial power spectrum is generally obtained by zero-padding arbitrarily the databox that contains the velocity field $U$ and averaging circularly with respect to the centre of the databox containing the Fourier transformed velocity field multiplied by the complex conjugated of its Fourier transform, i.e. by radially averaging with respect to the centre of the databox that contains the Cartesian power spectrum, which is given by

\begin{equation}
\mathcal{E}({\rm k_x,k_y})=\mathcal{F}\left\{ u({\rm x,y})\right\}_{\rm (k_x,k_y)}^*\mathcal{F}\left\{ u({\rm x,y})\right\}_{\rm (k_x,k_y)},\label{eqE}
\end{equation}

\noindent where $\mathcal{F}$ indicates the Fourier transform which maps from the real space extended/zero-padded grid $\mathcal{L}$ to the Fourier space extended grid $\mathcal{L}'$ when calculated through the FFT algorithm, * indicates the complex conjugated and $k_x$ and $k_y$ are the coordinates of Fourier space, i.e. the wavenumbers.

\subsection{Minimal zero-padding criterium for the correlation functions and power-spectrum }\label{sec:sec2.4}

Using the Wiener-Khinchin theorem [see e.gr. \citet{Champeney1973}] we get

\begin{equation}
B_{\square}({\rm x,y})=\mathcal{F}^{-1}\left\{\mathcal{E}({\rm k_x,k_y})\right\},\label{eqB}
\end{equation}

\noindent where $\mathcal{F}^{-1}$ indicates the inverse Fourier transform which maps from $\mathcal{L}'$ to $\mathcal{L}$, $B_{\square}({\rm x,y})$ is the Cartesian correlation function  normalised by the pixel area $a$ and the subindex $\square$ indicates normalisation by $a$.  In order to obtain the circular weighted correlation normalised by pixel area, $C_{\square}^{\rm w}(\delta r)$, we need to average over circles centred on the geometrical centre of the transformed databox that the inverse FFT yields and use our exact circular averaging scheme to ponder each pixel contribution. The previous is equivalent to making the transformation

\begin{equation}
B_{\square}({\rm x,y})\rightarrow^{\rm w} B_{\square}^{\rm w}(\delta r) \equiv C_{\square}^{\rm w}(\delta r),\label{eqBC}
\end{equation}

\noindent where $\rightarrow^{\rm w}$ indicates the transformation of weighted circular averaging described in the above paragraph. This last mapping allows to obtain $C_{\square}^{\rm w}(\delta r)$. Given that the field $U$ has finite extension there is a minimum amount of symmetrical zero-padding that needs to be added to calculate the correlation function up to $\delta r=D$ which is the displacement radius after which $C_{\square}^{\rm w}(\delta r)$ vanishes identically, i.e. it vanishes for all larger $\delta r$.\footnote{The higher-order correlation functions also vanish identically for $\delta r\ge D$} We will give this amount of zero-padding below.

In order to calculate  $C_{\square}^{\rm w}(\delta r)$ or higher order correlations up to $D$ one needs to zero-pad symmetrically the databox that contains $U$ up to obtaining an area in pixels of

\begin{equation}
\mathcal{A}_{\rm C}(D;\delta^+)=(2D+\delta^+)\times(2D+\delta^+) \mbox{pixels}^2; \label{eqZP3}
\end{equation}

\noindent  where $\delta^+=2$ pixels. The previous factor, $\delta^+$, takes into account the nature of the exact circular averaging and of the FFT as it provides a centre around which to carry out the circular averages and maintains the even parity of the sides of the databox for an efficient and accurate calculation of the FFT. In practice, after creating the array A of area $\mathcal{A}_{\rm C}(D;\delta^+)$ one has to centre the unpadded databox of area I$\times$J pixels$^2$ on it. The centre of the extended array is (i$_c$,j$_c$)=$(D+1,D+1)$.\footnote{Remember that we have assumed that the array indexes start at (0,0)} Then, to centre the unpadded databox in A, one must copy it setting its top or lower\footnote{This depends on the programming language that one has selected}  corner at the pixel with coordinates $(\delta_{\rm h},\delta_{\rm v})$ in A, where

\begin{equation}
\delta_{\rm h} = \left\lfloor \frac{2D+\delta^+-I}{2}\right\rfloor,
\end{equation}

\noindent

\begin{equation}
\delta_{\rm v} = \left\lfloor \frac{2D+\delta^+-J}{2}\right\rfloor.
\end{equation}

\noindent and $\lfloor\,\rfloor$ is the floor operator. Notice again that when the databox is completely filled with data $D$ is equal to the diagonal of the databox $L$. When more zero-padding that indicated by Eq. (\ref{eqZP3}) is added to calculate $C_{\square}^{\rm w}(\delta r)$ or higher order correlations, one simple obtains a trail of zeros for $\delta r>D$. 

The amount of zero-padding indicated by Eq.~(\ref{eqZP3}) is also the zero-padding required to calculate adequately the power-spectrum not only the correlation functions. Notice that when the power-spectrum is calculated with the amount of zero-padding indicated by Eq.~(\ref{eqZP3}) the second-order structure function obtained from this power spectrum using the Wiener-Khinchin theorem and Eq.~(\ref{eqS2w}) coincides exactly on the interval $\delta r=0$--$D$, up to machine numerical error, with the second-order structure function calculated using our real space algorithm, Eq.~(\ref{eqweighs_gcoords}). Because of this, the zero-padding indicated by Eq.~(\ref{eqZP3}) is the \emph{minimum optimal} zero-padding to calculate the power-spectrum or the correlation functions. However, the associated power-spectrum tends to exhibit small peaks locally and because of this we give preference to the zero-padding indicated in the next Subsection to calculate the power-spectrum as such amount of zero-padding is the \emph{maximum optimal}  to calculate $E(k)$. See also Section~\ref{sec:sec4} and Appendix~\ref{sec:secA} where we show the type of errors that occur when less than the zero padding indicated by Eq.~(\ref{eqZP3}) is added to calculate $E(k)$, $C_{\square}^{\rm w}(\delta r)$ and $S_{\rm 2}^{\rm w}(\delta r)$ through Fourier-transform-based methods.

\subsection{Maximum optimal zero-padding criterium for the power-spectrum}\label{sec:sec2.5}

Now, we will give a zero-padding criterium to extend symmetrically the unpadded databox that contains $U$ for an adequate calculation of the power-spectrum with optimal interpolation. Assume that the unpadded databox  that contains $U$ is filled completely with data and that the possibly filtered (masked) regions lay in the interior, i.e. without laying next the contour of the databox. Then, $D=L$. Thus, according to Eq. (\ref{eqZP1}) to be able to calculate $S_{\rm p,n}^{\rm w}(\delta r)$ through our real space algorithm up to the displacement $\delta r^*=D$ where the flat plateau begins,\footnote{This because the inertial range in some cases can extend up to $D$ and also because all variation of $S_{\rm p}^{\rm w}(\delta r)$ ends for $\delta r\ge D$ where it becomes one and the perfectly flat plateau begins} one needs to zero-pad symmetrically the original databox up to obtaining an area 

\begin{equation}
\mathcal{A}_{\rm ZP}(L;\rm I, J)=(\text{I}+4L)\times(\text{J}+4L).\label{eqZP4}
\end{equation}

Our calculation criterium is that to compute the power-spectrum through the FFT, Eq. (\ref{eqE}), without any bleeding or aliasing, at least the amount of symmetrical zero-padding indicated by Eq. (\ref{eqZP4}) must be added to the original  databox containing $U$, provided it is completely filled with data. In other words, to calculate the power-spectrum optimally, one must use a databox extended to the same area than one that would used to calculate $S_{\rm p}^{\rm w}(\delta r)$ up to $\delta r=\delta r^*=D=L$ through Eq. (\ref{eqweighs_gcoords}) or Eq.~(\ref{eqweighs_gcoords_odd}). The logic behind this criterium is that this databox extension provides what is required to calculate $S_{2}^{\rm w}(\delta r)$ up to $D$ through our real space algorithm, Eq.~(\ref{eqweighs_gcoords}), \emph{and that this $S_{2}^{\rm w}(\delta r)$ coincides exactly up to $D$ with the $S_{2}^{\rm w}(\delta r)$ calculated through the Fourier-transform-based algorithm, as demonstrated in Paper I. Thus, because of this, the power-spectrum associated to the databox zero-padded up to an area $\mathcal{A}_{\rm ZP}(L;\rm I, J)$, Eq.~(\ref{eqZP4}), must be the \emph{maximum optimal} power-spectrum, otherwise the second-order structure functions calculated through both methods would not coincide exactly up to machine numerical error.}

Because of one needs to average circularly around a centre and because the FFT usually requires an even number of pixels on each side of the databox to be transformed for a more precise and fast calculation, one may not be able to always use directly Eq.~(\ref{eqZP4}) to zero-pad to calculate the power-spectrum optimally. Instead, one needs to use the slightly modify formula

\begin{equation}
\mathcal{A}_{\rm ZP}(L;\rm I, J)=(\text{I}+4L+\delta_{\rm I}^{+})\times(\text{J}+4L+\delta_{\rm J}^{+});\label{eqZP5}
\end{equation}

\noindent where $\delta_{\rm I}^{+}=2$ pixels if I is even and $\delta_{\rm I}^{+}=1$ pixel if I is odd. Something analogue applies to $\delta_{\rm J}^{+}$, $\delta_{\rm J}^{+}=2$ pixels if J is even and $\delta_{\rm J}^{+}=1$ pixel if J is odd.

Thus, after calculating the power-spectrum through Equation~(\ref{eqE}) one needs to average radially around the centre of the zero-padded databox to obtain the radial power-spectrum, $E(k)$, where $k$ is the radial wavenumber. We will asumme that the databox have been zero-padded as indicated in the Eq.~(\ref{eqZP5}).  Hence, the centre of the databox around which the circular averages will be calculated has pixel coordinates $(i_{\rm c},j_{\rm c})$ given by

\begin{equation}
i_{\rm c}=2L+\left\lfloor\frac{I+\delta_{\rm I}^{+}}{2}\right\rfloor
\end{equation}

\noindent and

\begin{equation}
j_{\rm c}=2L+\left\lfloor\frac{J+\delta_{\rm J}^{+}}{2}\right\rfloor.
\end{equation}

When the databox is not filled up completely and there is a large amount of zeros surrounding non-vanishing data, one needs first to reduce the original databox to the tightest rectangular databox containing the non-vanishing data. Let's the sides of this reduced databox be I'$\times$J'.  Then the diagonal of the reduced databox is $L'=\lceil\sqrt{\rm I'^2+J'^2}\rceil$ and one needs to zero-pad using Eq~(\ref{eqZP5}) using the primed parameters (I',J' and $L$') instead of the un-primed ones.

\section{Structure functions up to the 8-th order}\label{sec:sec3}

Sometimes it is possible to diagnose the turbulence in a three-dimensional region through its associated 2D LOS centroid velocity map. \citet{Brunt2004} have determined through numerical simulations that if driven turbulence is ongoing in a three-dimensional region the exponent $\alpha_2$ of the second-order structure function corresponding to the LOS centroid velocity map approximates closely the exponent of the second-order structure function of the associated three-dimensional region. The turbulence in the Huygens Region may be driven, at least partially, due to the presence of young stars, proto-stars and Herbig-Haro objects which may drive the turbulence in the region through winds, jets and radiation. Furthermore, another indicator that the turbulence in the Huygens Region may be driven is that it is an H II region whose LOS centroid velocity PDF follows a shape similar to that of forced solenoidal turbulence, see Figure~\ref{fig:fig2}(b). On the other hand, \citet{Miville2003}, have determined through simulations that for optically thin lines the spectral index $\beta$ of the power-spectrum of the 2D LOS centroid velocity map is the same than the spectral index $\beta_{\rm 3D}$ of the associated three-dimensional field. Since the \emph{MUSE} H$\alpha$ LOS centroid velocity map that we are analysing was derived from optically thin lines since the nebular gas in the Huygens Region is not dense enough to have significant self-absorption (Weilbacher --private communication), then the power-spectrum obtained from the LOS centroid velocity map is very probably proportional to the one associated to the three-dimensional velocity field of the region, see also \citet{ZuHone2016}.  The higher-order transverse structure functions are probably also similar to those of the associated three-dimensional field. We will assume that this is the case, i.e. that the exponents of the power-laws associated to both the transverse structure functions and the power-spectrum corresponding to the H$\alpha$ LOS centroid velocity map coincide with the exponents associated to the three-dimensional velocity field of the Huygens region.

\begin{figure*}
	\includegraphics{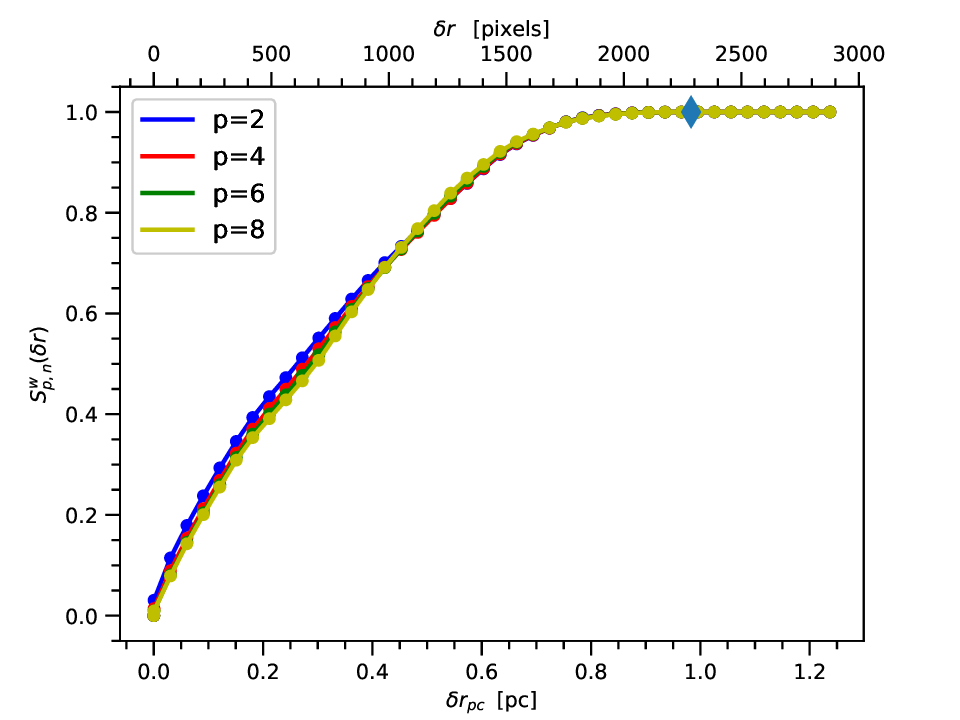}
	\caption{Normalised even order structure functions of orders 2,4,6 and 8. The x-axis tickmarks are given in parsecs and pixels. All the presented structure functions are very similar to each other, only very small deviations are observed for $\delta r$ between 0.129--0.430 pc (300--1000 pixels).The blue diamond marks the point when the flat plateau begins, i.e, the value corresponding to $\delta r=D=2286$  pixels (0.98298-pc) after which the normalised structure functions take exacly the value of 1. The inertial range determined through fitting the second-order function to a power-law corresponds to the interval $\delta r=0.0301$--0.645 pc (70--1500 pixels). The structure functions of all orders coincide very closely after a separation distance $\delta r=0.430$-pc (1000 pixels). Because of this, we fitted the structure functions to power-laws on the interval $\delta r=0.129$--0.430-pc (300--1000 pixels) in order to measure the small deviations that occur on this sub-interval of the inertial range. The result of the fittings are given in Table~\ref{tab:tab2}.}\label{fig:fig5}
\end{figure*}

We calculated the normalised structure functions of the extended turbulent field $U$ up to the 8th order using Eqs.~(\ref{eqweighs_norm})~and~(\ref{eqweighs_norm_odd}). The normalised even-order structure functions are presented in Figure~\ref{fig:fig5} and the normalised odd-order structure functions are presented in Figure~\ref{fig:fig7}. Given that our circular averaging algorithm is based on exact computational geometry derivations and our calculation through Eqs.~(\ref{eqweighs_norm})~and~(\ref{eqweighs_norm_odd}) was exhaustive, the presented structure functions are exact up to machine rounding and truncation errors.

The x-axes of Figures~\ref{fig:fig5} and \ref{fig:fig7} are given in parsecs and pixels. In order to convert $\delta r$ from pixels to parsecs one needs to multiply the value of $\delta r$ in pixels by the spatial length $l=0.00043$-pc covered by one pixel of the instrumentation of \emph{MUSE} for the data we are analysing, see \citet{Weilbacher2015}. We label the transverse separation distance in parsecs as $\delta r_{\rm pc}$ which is given thus by

\begin{equation}
	\delta r_{\rm pc}=l\delta r=0.00043\delta r\,\mbox{[pc]}\label{eqdeltarpc}.
\end{equation}

Notice that as discussed in Subsection~\ref{sec:sec2.2}, all normalised structure functions converge to 1 after a separation distance D=2286 pixels, which is slightly less than the diagonal of the databox $L=2302$ pixels because there is missing data close to the borders of the databox (particularly next to the corners) because some pixels there are zero-valued in the original (non-filtered) LOS centroid velocity map. The normalisation by $k_{\rm p}$ allows to compare the even-order structure functions on an equal basis, which allows a quickly visual determination of their differences and similarities without using logarithmic scale (where they are usually shifted vertically from each other)  and before using regression (fitting) to determine the best power-laws that can be adjusted on the interval corresponding to the inertial range in order to make an analytical diagnose. For the turbulent field $U$ of the Huygens Region, one can see that both the normalised higher-order even  and odd transverse structure functions are very similar to the normalised second-order transverse structure function.

\begin{figure}
	\centering
	\includegraphics[width=\columnwidth]{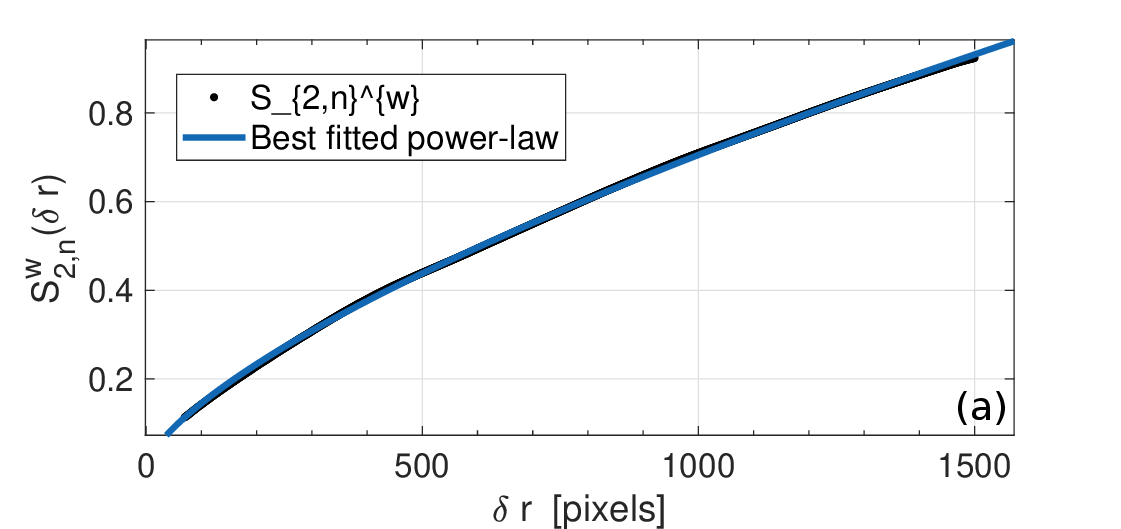}\\
	\includegraphics[width=\columnwidth]{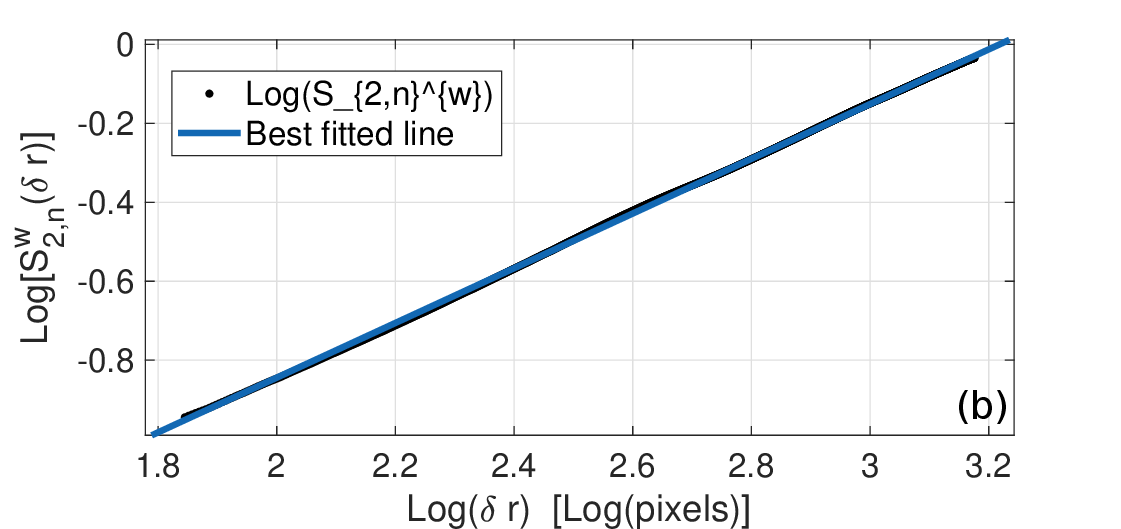}\\
	\caption{Panel (a): inertial range determined by fitting $S_{2}^{\rm w}(\delta r)$ to a power-law in the interval $\delta r$=70--1500 pixels (0.0301--0.645 pc). We find an exponent $\alpha_2=0.6868$ with $R^2=0.9998$. Panel (b): linear fitting of the inertial range using log-log scale. We find an exponent $\alpha_2=0.6929$ with $R^2=0.9998$.}
	\label{fig:fig6}
\end{figure}

\begin{figure*}
	\centering
	\includegraphics{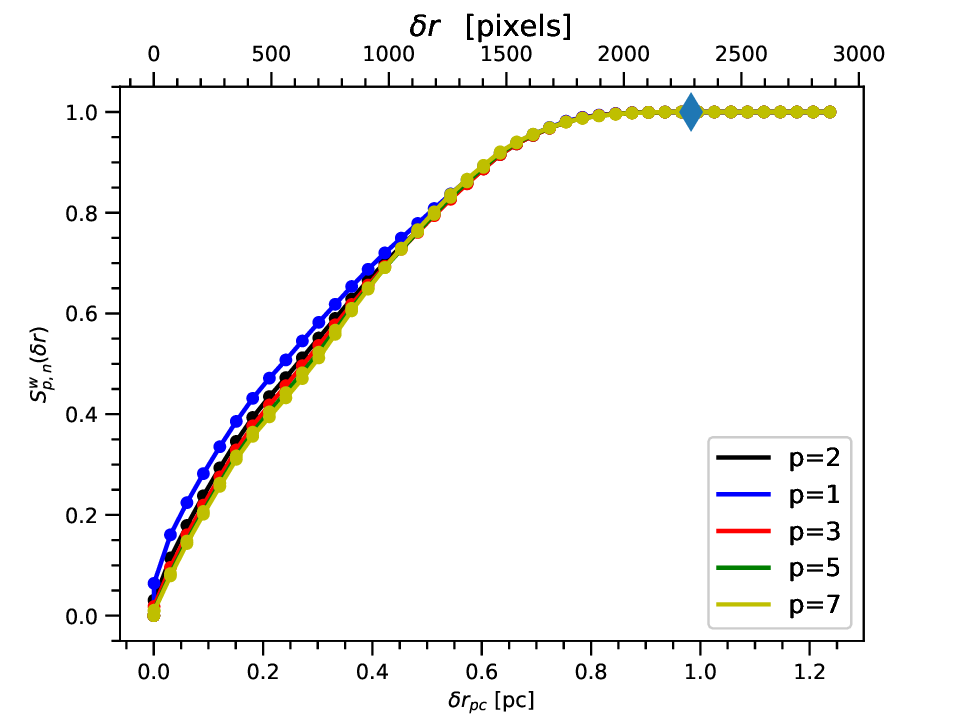}
	\caption{Normalised odd-order structure functions for p=1,3,5 and 7. The second-order structure function is traced in black as a reference.  The blue diamond has the same meaning that in Figure~\ref{fig:fig5}. We find that for p>1 the transverse structure functions are very similar to $S_{2}^{\rm w}(\delta r)$.}\label{fig:fig7}
\end{figure*}

 In Paper I, the inertial range of the second-order transverse structure function was determined to correspond to the interval $\delta r=300$--1500 pixels (0.129--0.645-pc) where $S_{2}^{\rm w}(\delta r)$ follows a power-law with exponent $\alpha=0.6824$. However, we found this inertial range without information from the power-spectrum, Section~\ref{sec:sec4}, and we were too strict in setting the lower limit of the inertial range to avoid the initial interval of $\delta r$ where $S_{2,n}^w$ has a behaviour that is different from a power-law. Now with information from the power-spectrum, Figure~\ref{fig:fig9}, we find that the inertial range actually\footnote{We were very careful this time in avoiding as much as possible the  initial profile of $S_{2,n}^w$ that does not follow a power-law using strict real space regression and taking into account also information from the power-spectrum} extends from $\delta r=70$ pixels (0.0301-pc) to 1500 pixels (0.645-pc) where it has a power-law behaviour with exponent $\alpha_2=0.6868$ if a power-law is fitted directly or $\alpha_2=0.6929$ if a line is fitted in log-log scale; both exponents are very close to the value of 0.6824 found in Paper I. The plots corresponding to the previous fittings are shown in Figure~\ref{fig:fig6}; one can see that the power-law fittings are very precise, they both have a multiple determination coefficient of $R^2=0.9998$. The higher-order structure functions seem to follow to a high degree of approximation also this power-law behaviour with very small deviations only in a subinterval of the inertial range which corresponds to $\delta r=300$--1000 pixels ($\delta r_{\rm pc}=0.129$--0.430 pc). The similarity between all order structure functions is remarkable. We will analyse and find the physical meaning of this similarity in Sections~\ref{sec:sec5.1}~and~\ref{sec:sec5.2}.

Noticeably, we do not find any scaling similar to that proposed by \citet{She1994} --SL94 hereafter-- among the higher-order transverse structure functions. The SL94 theory predicts that the velocity structure functions scale as power laws with exponents $\zeta_{\rm p}=p/9+2[1-(2/3)^{p/3}]$ in the inertial range.  This scaling is often assumed or found in the literature. Here, in turn, we have found only very small deviations among the higher-order transverse structure functions up to order 8th after normalisation; this implies that the higher-order transverse structure functions are almost proportional to the second-order transverse structure function, i.e. that they differ almost only by a constant multiplicative factor. This contradicts also the often invoked scalings of the higher-order structure functions as a power of $S_{2}(\delta r)$ or $S_{3}(\delta r)$, i.e.  scalings of the form $S_{\rm p}\propto S_{\rm 2}(\delta r)^{\gamma_p}$  or $S_{\rm p}\propto S_{\rm 3}(\delta r)^{\gamma_p}$  where $\gamma_{\rm p}$ are often exponents inspired on the SL94 theory. In order to evaluate analytically how much similar the higher-order structure functions presented in Figures~\ref{fig:fig5}~and~\ref{fig:fig7} are, we fitted them to power-laws in the interval $\delta r=300$--1000 pixels (0.129--0.430 pc) where they differ the most. The results are given in Table~\ref{tab:tab2}, where we also give the exponent predicted by the SL94 theory. One can see that the exponents $\alpha_{\rm p}$ of the adjusted power-laws are very similar, with a maximum deviation of 0.0989 between $\alpha_2$ and $\alpha_8$.\footnote{Ignoring $\alpha_1$ since $S_{1}^{\rm w}(\delta r)$ is the transverse structure function that deviates the from $S_{2}^{\rm w}(\delta r)$ because of its amplitude $A_{\rm p}$ and slightly small exponent.} Notice however, that besides the exponents $\alpha_p$, an amplitude $A_{\rm p}$ was also result of the fitting of power-laws of the form $A_{\rm p}\delta r^{\alpha_{\rm p}}$. The amplitudes $A_{\rm p}$ found make the higher-order transverse structure functions similar to the second-order transverse structure function despite small differences in their exponents, $\alpha_{\rm p}$. On the other hand the SL94 theory predicts approximately correctly only the exponent $\alpha_2$ corresponding to the the second-order transverse structure function and predicts and exponent too small for the first order transverse structure function and exponents too large for the higher-order transverse structure functions. Scalings \a'a la SL94 have been found by \citet{Padoan2003} [see also \citet{Boldyrev2002}] for intensity maps of the $J=1-0^{13}$CO transition in the Taurus and Perseus molecular clouds, thus the scaling proposed by SL94 does not seem to exclusively apply for three-dimensional fields. Scalings \a'a la SL94 have been also used or reported by many other authors. However, it seems that for the case of the Huygens Region the situation is completely different. In Section~\ref{sec:sec5} we will demonstrate that the scaling that we found for the higher-order structure functions of the Huygens Region implies a particular type of homogeneity that we will find and analyse through a simple mathematical model.

\begin{table}
	\centering
	\caption{Fitting parameters of the power-laws to which the normalised transverse structure functions were fitted on the interval 300-1000 pixels (0.129--0.430-pc). The power laws are of the form $S_{\rm p}^{\rm w}(\delta r)=$ A$_{\rm p} \delta r^{\alpha_{\rm p}}$. The multiple determination coefficient is $R^2$. We also give the exponents $\zeta_{\rm p}$ predicted by the \citet{She1994} theory.}
	\begin{tabular}{lcccc} % four columns, alignment for each
		\hline
		p   & A$_{\rm p}$& $\alpha_{\rm p}$ & R$^2$    & $\zeta_{\rm p}$   \\		\hline
	    1   & 0.010600   & 0.6064           & 0.9997   & 0.3640           \\
		2   & 0.006136   & 0.6868           & 0.9998   & 0.6959       \\
		3   & 0.004200   & 0.7281           & 0.9995   & 1.0000       \\                                
		4   & 0.004100   & 0.7446           & 0.9993   & 1.2797           \\
		5   & 0.003600   & 0.7605           & 0.9989   & 1.5380          \\
		6   & 0.003400   & 0.7696           & 0.9985   & 1.7778           \\
		7   & 0.003100   & 0.7803           & 0.9976   & 2.0013         \\
		8   & 0.003000   & 0.7857           & 0.9966   & 2.2105           \\
		\hline
	\end{tabular}\label{tab:tab2}
\end{table}

 The similarity of the higher-order transverse structure functions to $S_{2}^{\rm w}(\delta r)$ is alike to the similarity that occurs in Burgers turbulence also called burgulence. In the case of Burgers turbulence, for integer p$\ge$1, all the transverse structure functions are proportional to $\delta r$ with distinct proportionality coefficientens, i.e. in the case of burgulence $S_{\rm p}\propto\delta r$, see \citet{Frisch1995} and \citet{Xie2021} and see also \citet{Schmidt2008} and \citet{Federrath2013}. The scaling of burgulence for integer p$\ge 1$ is due to the dominance of shocks over other structures in the flow. From similar considerations, \citet{Boldyrev2002} proposed an hybrid Kolmogorov-Burgers model in which $S_{2}(\delta r)\propto \delta r^{0.74\mbox{--}0.76}$. In his model, the dissipative structures are quasi-one-dimensional shocks. Given that the transverse structure functions of the Huygens Region have an analogue invariant scaling than that of Burgers turbulence, and given that their exponents in the inertial range are close to the Kolmogorov exponent and the exponents found by \citet{Boldyrev2002} for $S_2(\delta r)$, see Table~\ref{tab:tab2}, it is possible that the turbulence in the Huygens Region may be dominated by shocks, or at least, it seems that shocks might play an important role in establishing the spatial configuration of the velocity field of Huygens Region. In Section~\ref{sec:sec5.1}, we will demonstrate that the spatial configuration of the LOS centroid velocity field of the Huygens Region possesses a particular type of homogeneity. Given that the LOS centroid velocity field  of the Huygens Region is representative of the corresponding three-dimensional velocity field, see \citet{Brunt2004} and \citet{Miville2003}, it is plausible to consider that the three-dimensional field of the Huygens Region also posses such type of homogeneity.

For comparison purposes, we calculated the normalised transverse structure functions corresponding to the full PDF shown in Figure~\ref{fig:fig2}. This PDF corresponds to the raw data filtered between -9$\sigma$--5$\sigma$ limits and thus contains the data corresponding to the broad left tail of the PDF which corresponds to the quiescent region, i.e. the quiescent region was not filtered out this time. We  find that the higher-order transverse structure function deviate even less from each other than those shown in Figures~\ref{fig:fig5}~and~\ref{fig:fig7}, however the deviations are more irregular in the sense that the structure functions do not follow now a decreasing order with increasing p after normalisation by $k_{\rm p}$. This is due  to the fact that quiescent region was taken into account. Withal, this result demonstrates that the fact that the higher-order even transverse functions are almost proportional to $S_{2}^{\rm w}(\delta r)$ is an intrinsic property of the LOS centroid velocity field of the Huygens Region and not a product of the filtering of the quiescent region. The exponents in the inertial range of the transverse structure functions when the quiescent region is not filtered, particularly that of $S_{2}^{\rm w}(\delta r)$, are closer to the exponents for $S_2(\delta r)$ found by \citet{Boldyrev2002} for hybrid Kolmogorov-Burgers turbulence, 0.74--0.76, see Paper I.

\begin{figure*}
	\includegraphics{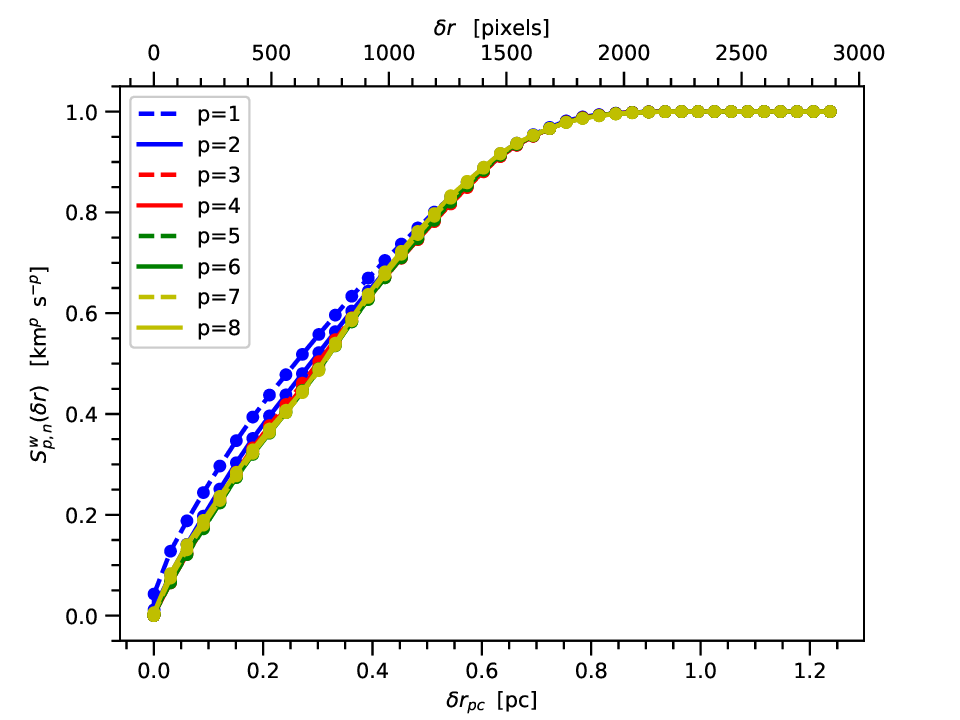}
	\caption{Normalised transverse structure functions of orders from 1st to 8th for the data corresponding to the PDF shown in Figure~\ref{fig:fig2}. The PDF contains the data of the map of the Huygens Region within -9$\sigma$--5$\sigma$ limits. The x-axis tickmarks are given in pc and pixels. The blue diamond has the same meaning than in Figure~\ref{fig:fig5}. The associated structure functions deviate less than each other than those presented in Figures~\ref{fig:fig5}~and~\ref{fig:fig7}, however the deviations are more irregular because the quiescent region was not filtered out from the LOS centroid velocity map.}\label{fig:fig8}
\end{figure*}

\section{Power spectrum}\label{sec:sec4}

In this Section we present the power-spectrum of the turbulent field $U$ of the Huygens Region. We do not calculate it using averaging over bands as is typically done in the literature, instead we use our exact circular averaging algorithm based on computational geometry through which we can obtain the exact power-spectrum.  We will  discuss the physics behind the obtained  power-spectrum in Section~\ref{sec:sec5.3}. For simplicity, we will use the spatial frequency number $n$ and the associated separation distance $\delta r$ in pixels instead of the wavenumber for our representation of the power-spectrum as this way the plots are easier to read and this does not affect the exponent of the power-laws to which the power-spectrum will be fitted. The relationship between $n$  with $\delta r$ in pixels is

\begin{equation}
\delta r=\frac{L}{n}=\frac{2302\,\mbox{[pixels]}}{n}.\label{eqdeltar-n}
\end{equation}

\noindent For converting $\delta r$ to parsecs one needs to use Eq.~(\ref{eqdeltarpc}).

\begin{landscape}
	\begin{figure}
		\centering
		\includegraphics[scale=0.45]{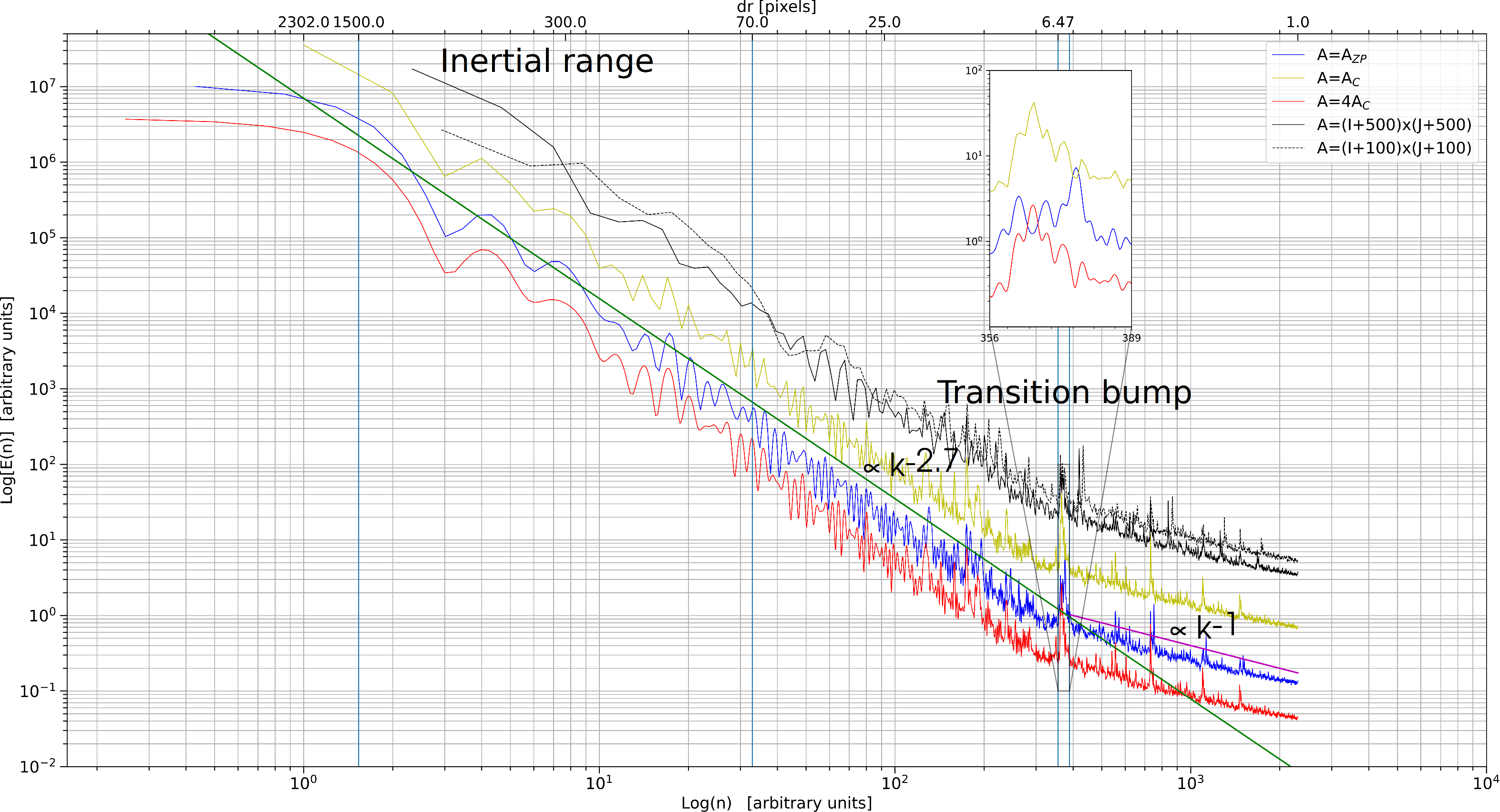}\\
		\caption{Power-spectra obtained using different amounts of zero-padding. The different curves were displaced vertically for a better visualisation. The area of the extended databox is $A$ and the different $A$'s used are indicated in Table~\ref{tab:tab2}. The yellow curve uses the exact amount of zero-padding needed to obtain $C_{\square}^{\rm w}(\delta r)$ up to $\delta r=D\approx L$ where it vanishes, see Eq. (\ref{eqZP3}). The blue curve uses our maximum optimal criterium indicated by Eq. (\ref{eqZP5}), the corresponding power-spectrum has better spectral resolution and evince a smoother behaviour. The red curve corresponds to much more zero-padding, the area of its extended databox is four times the area of the databox used to obtain the yellow curve. The black lines show degraded spectra due to lack of sufficient zero-padding. The green line represents a power-law with exponent $\beta=-2.7$ which corresponds to Kraichnan turbulence and the magenta line represents a power-law with exponent $\beta=-1$ which implies viscous dissipative processes occurring at the corresponding small spatial scales. The blue curve is well represented by these power-laws. The inset box shows the  characteristic transition bump between the two power-laws. The inertial range determined through the second-order structure function $S_{2}^{\rm w}$ is indicated between blue vertical lines.}
		\label{fig:fig9}
	\end{figure}
\end{landscape}

We calculate the power-spectrum through Eq. (\ref{eqE}) with different amounts of zero-padding to test our zero-padding criteria given in Subsections~\ref{sec:sec2.4}~and~\ref{sec:sec2.5}, Eqs.~(\ref{eqZP3}) and (\ref{eqZP5}). We show the associated power-spectra in Figure~\ref{fig:fig9} and in Table~\ref{tab:tab3} we give the amounts of zero-padding used. We find that the minimal optimal zero-padding criterium to calculate the correlation function(s) and the power-spectrum given in Subsection~\ref{sec:sec2.4}, Eq.~(\ref{eqZP3}), yields a good quality power-spectrum that however exhibits small local peaks. On the other hand, we find that the maximum optimal criterium given in Subsection~\ref{sec:sec2.5}, Eq.~(\ref{eqZP5}), yields a smoother power-spectrum because its better spectral resolution consequence of the larger amount of zero-padding used. A power-spectrum  was calculated with even more zero-padding, with 4 times the zero padding indicated by Eq.~(\ref{eqZP3}), this power-spectrum has naturally a better spectral resolution; however the improvement is not significant, which demonstrates that the zero-padding criteria given by Eqs.~(\ref{eqZP3})~and~(\ref{eqZP5}) are optimal. In order to demonstrate this, we also calculated the power-spectrum using less zero-padding than indicated by Eq.~(\ref{eqZP3}): we calculated  the power-spectrum also with a symmetrical zero-padding of 100 and 500 pixels only. Clearly, the power-spectrum calculated with only 100 pixels of symmetrical zero-padding is degraded and deviates significantly from the truth power-spectrum at intermediate spatial frequencies and does not cover the spatial frequencies corresponding to the inertial range fully. The power-spectrum calculated with 500 pixels of symmetrical zero-padding shows less deviations but it is still degraded and does not covers fully neither all the spatial frequency numbers corresponding to the inertial range. This shows the importance of using the optimal amount of symmetrical zero-padding to calculate the power-spectrum which is given by Eqs.~(\ref{eqZP3}) and (\ref{eqZP5}).

\begin{table}
	\centering
	\caption{Sizes of the databoxes used in the calculation of the power-spectra shown in Figure~\ref{fig:fig9}. The sides of the unpadded databox are I=1766 pixels and J=1476 pixels. The diagonal of the unpadded databox is $L=2302$ pixels and $\delta_{\rm I}^+=\delta_{\rm J}^+=2$ pixels.}\label{tab:tab3}
	\begin{tabular}{lccc} % four columns, alignment for each
		\hline
		Spectrum & Size in pixels$^2$ of the databox                                  & Line colour \\
		         & after zero-padding it symmetrically                                   &             \\
		\hline
		1        & $(2L+\delta^+)\times(2L+\delta^+)$                                 & Yellow\\                                 
		2        & $({\rm I}+4L+\delta_{\rm I}^+)\times({\rm J}+4L+\delta_{\rm J}^+)$ & Blue  \\
		3        & $(8L+\delta^+)\times(8L+\delta^+)$                                 & Red   \\
		4        & (I+500)$\times$(J+500)                                             & Black \\
		5        & (I+100)$\times$(J+100)                                             & Black (dashed)\\
		\hline
	\end{tabular}
\end{table}

\begin{figure}
	\centering
	\includegraphics[width=\columnwidth]{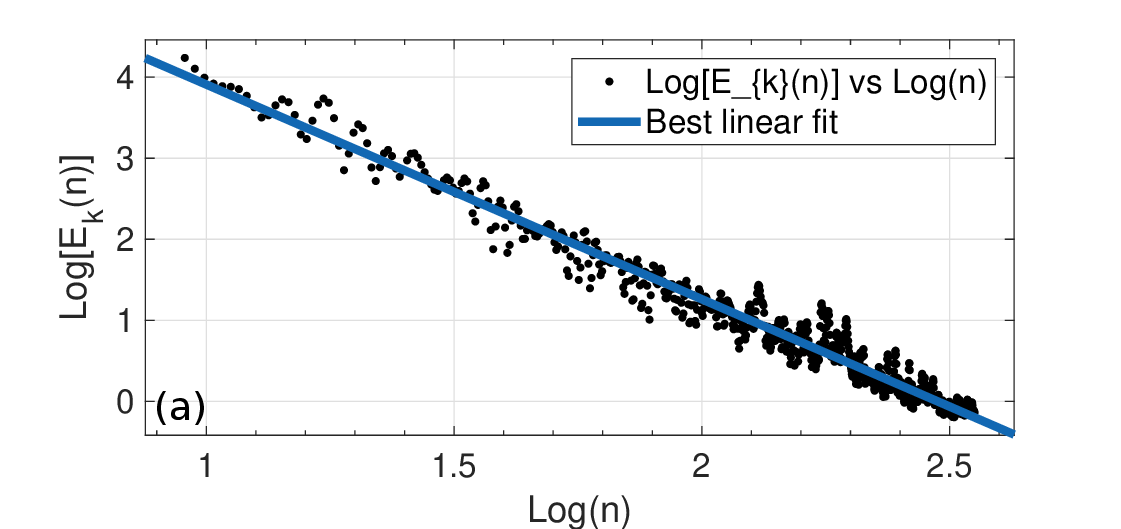}\\
	\includegraphics[width=\columnwidth]{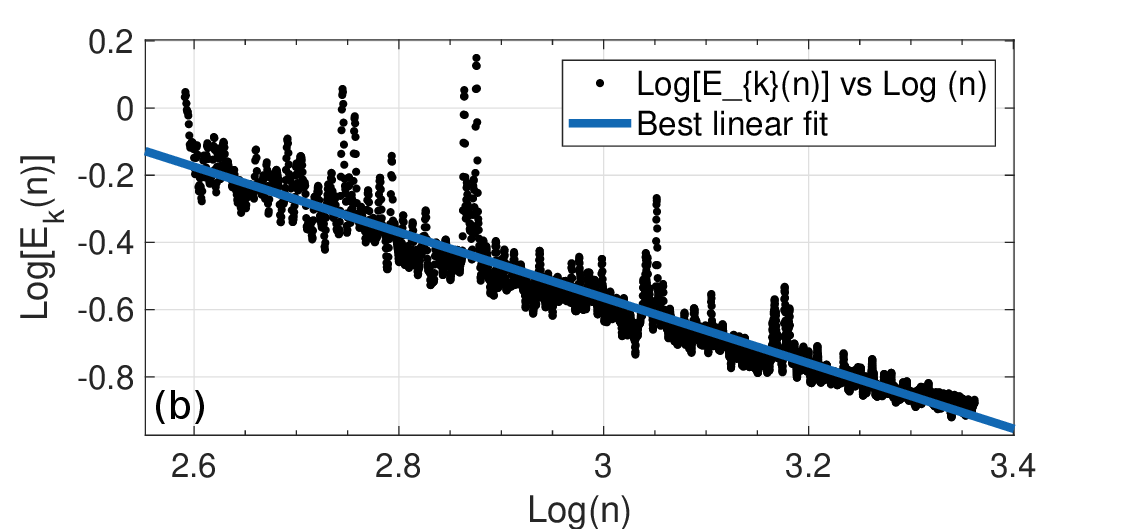}\\
	\caption{Regression plots for the tail of the power-spectrum of the turbulent field $U$ of the Huygens Region. Panel (a): regression plot for spatial frequencies $n=9$--355 where a power-law with exponent $\beta=-2.6500$ is found to adjust with $R^2=0.9678$. Panel (b): regression plot for spatial frequencies $n=390$--2302 where a power-law with exponent $\beta=-0.9747$ is found to adjust with $R^2=0.9268$. }
	\label{fig:fig10}
\end{figure}

We find that the power-spectrum exhibits an initial plateau that declines from $n\approx 2$--4. Later it exhibits two intermediate size bumps between $n=5$--8 and after that it exhibits a long tail that extends over all the inertial range and reaches spatial frequencies of up to $n=355$. After  that there is a noticeably spectral bump that covers an spectral range between $n=356$--389 (which corresponds to an spatial scale of $\delta r\sim 6$ pixels or $\delta r_{\rm pc}=0.00258$-pc) and after which the slope of the power-spectrum changes abruptly. These two regions of the long tail separated by the spectral bump can be fitted to two power-laws, Figure~\ref{fig:fig9}. We performed least-squares regression to fit to power-laws the two relevant segments of the long tail and we give the results of the regression in Table~\ref{tab:tab4}.  The best fitting lines are shown in Figure~\ref{fig:fig10}. We find that the initial side of the tail of the power-spectrum can be fitted to a power-law with exponent $\beta_1=-2.6500$ between $n=9-355$ which corresponds to small and intermediate spatial frequencies which have associated spatial scales of $\delta r\sim 6$--255 pixels or $\delta r_{\rm pc}\sim 0.00258$--0.10965 pc. It was not possible to fit a reliable power-law for $n<9$ because of the slope of the plateau and the two intermediate size bumps at those frequencies. After the transition bump the power-spectrum  adopts a second power-law with exponent $\beta=-0.9747$ . We find that this last exponent corresponds to $n=390$--2302 which in turn corresponds to spatial scales of $\delta r\sim 1$--5 pixels or $\delta r_{\rm pc}\sim 0.00043$--0.00215 pc.

\begin{table}
	\centering
	\caption{Fitting parameters of the power-laws to which the tails of the power-spectrum of the field $U$ of the Huygens Region were fitted. The power-laws are of the form $E(n)\propto n^{\beta}$. The variable $\delta n$ indicates the spatial frequency interval that corresponds to each power-law. The multiple determination coefficient is $R^2$. We also give 95\% confidence bounds for the exponents $\beta$. The approximate extension of the bump that separates the two power-laws is also given.}
	\label{tab:example_table}
	\begin{tabular}{lccccc} % four columns, alignment for each
		\hline
		Tail         & $\beta$    & $\delta n$            & $R^2$  & 95\% confidence  \\
		             &            & [1/pixel]             &        & bounds for $\beta$\\
		\hline
		initial side & -2.6500    & 9 -- 355              & 0.9678 & -2.6830 -- -2.6170\\                                 
		Bump         &            & 356 -- 389\\
		end side     & -0.9747    & 390 -- 2302           & 0.9268 & -0.9827 -- -0.9666\\
		\hline
	\end{tabular}\label{tab:tab4}
\end{table}

We remark that for the spatial-frequency interval between  $n=(L/1500)$--8 ($n=1.5347$--8  with $L=2302$ pixels) the power-spectrum does not follow reliably a power-law. This makes evident the relevance of $S_{2}^{\rm w}(\delta r)$ for the determination of the inertial range as \citet{Kolmogorov1991} determined it, since in general $S_{2}^{\rm w}(\delta r)$ is much more stable than the power-spectrum. However, by visual inspection, one can roughly conclude that the tail with exponent $\beta=-2.6500$ extends from $\delta r=70$--1500 pixels (0.0301--0.6450 pc), which corresponds to the inertial range. For smaller $\delta r$ (larger $n$) the power spectrum becomes peaky/noisy and presents tight wave packets. For such noisy region the second-order structure function does not follow a power-law and because of that we consider that the power-law with exponent $\beta=-2.6500$ is representative only on the interval $\delta r=70$--1500 pixels (0.0301--0.6450 pc).

\section{Discussion}\label{sec:sec5}

In this Section we interpret the meaning of the similarity between the normalised transverse structure functions and also find an interpretation for the power-laws to which the tail of the power-spectrum was fitted. We do this in terms of simple mathematical models.

\subsection{Similarity of the even-order structure functions to $S_{\rm 2,n}^{\rm w}(\delta r)$ after normalisation: homogeneity}\label{sec:sec5.1}

In order to construct a mathematical model to interpret the similarity of the even-order structure functions after normalisation by $k_{\rm p}$, Figures~\ref{fig:fig5}~and~\ref{fig:fig8}, we first expand the binomial in Eq.~(\ref{eqweighs_gcoords}) to obtain

\begin{equation}
S_{\rm p}^{\rm w}(\delta r) =\sum_{\rm m,n}^{\rm M'',N''}\sum_{\theta=0}^{2\pi} {\rm w}(\theta;\delta r) \sum_{{\rm i}=0}^{\rm p} (-1)^{\rm i} \binom{\rm p}{{\rm i}} u({\bf r})^{\rm p-i} u({\bf r+}\delta {\bf r})^{\rm i}. \label{eqgenbraexp1}  
\end{equation}

\noindent Extracting the first and last term from the inner summation containing the binomial term we get

\begin{equation}
S_{\rm p}^{\rm w}(\delta r) =\sum_{\rm m,n}^{\rm M'',N''}\sum_{\theta=0}^{2\pi} {\rm w}(\theta;\delta r) \left[u({\bf r})^{\rm p} + u({\bf r}+\delta {\bf r})^{\rm p}\right] + \label{eqgenbraexp2}  
\end{equation}

\begin{equation*}
\sum_{\rm m,n}^{\rm M'',N''}\sum_{\theta=0}^{2\pi} {\rm w}(\theta;\delta r) \sum_{{\rm i}=1}^{\rm p-1} (-1)^{\rm i} \binom{\rm p}{{\rm i}} u({\bf r})^{\rm p-i} u({\bf r+}\delta {\bf r})^{\rm i}.  
\end{equation*}

In Paper II, we demonstrate analytically that each of the first two terms in the previous Equation is equal to $\sum_{{\rm (i,j)}\in\mathcal{L}_0} u({\rm i,j})^{\rm p}$; then, we can write the previous Equation as

\begin{equation}
S_{\rm p}^{\rm w}(\delta r) =k_{\rm p} +\mathcal{C}_{\rm p}^{\rm w}(\delta r),\label{eqkpCp}
\end{equation}

\noindent where $k_{\rm p}$ is given by Eq.~(\ref{eqkp}) and $\mathcal{C}_{\rm p}^{\rm w}(\delta r)$ is given by\footnote{Notice that for a fast calculation, the indices m,n,M'',N'' can be set this time to 0,0,I-1,J-1 because of the nature of the correlation function \emph{provided the databox is zero-padded using Eq.~(\ref{eqZP3}) and the circular averages are fully carried out even if they cross the border of the I$\times$J databox}. For simplicity we keep the indexes in the respective summation symbol as m,n,M'',N'' as this does not affect the results nor the conclusion of our present analysis}

\begin{equation}
\mathcal{C}_{\rm p}^{\rm w}(\delta r) =\sum_{\rm m,n}^{\rm M'',N''}\sum_{\theta=0}^{2\pi} {\rm w}(\theta;\delta r) \sum_{{\rm i}=1}^{\rm p-1} (-1)^{\rm i} \binom{\rm p}{{\rm i}} u({\bf r})^{\rm p-i} u({\bf r+}\delta {\bf r})^{\rm i}. \label{eqCp}  
\end{equation}

\noindent Notice that $\mathcal{C}_{\rm p}^{\rm w}(\delta r)$ can be interpreted as a p-th order correlation function composed of the alternating sum of binomially weighted p-th order correlation functions of the form $C_{\rm p-i,i}^{\rm w}(\delta r)$, where

\begin{equation}
	C_{\rm p-i,i}^{\rm w}(\delta r)=\sum_{\rm m,n}^{\rm M'',N''}\sum_{\theta=0}^{2\pi} {\rm w}(\theta;\delta r)   u({\bf r})^{\rm p-i} u({\bf r+}\delta {\bf r})^{\rm i}. \label{eqCij}
\end{equation}

Now, in order to propose a model to explain the similarity between the even-order structure functions, we remind the reader of the concept of an homogeneous operator. An operator (or function) $\mathcal{O}(x)$ is homogeneous if

\begin{equation}
	\mathcal{O}(tx)=t^{\rm k}\mathcal{O}(x),
\end{equation}

\noindent where $t$ is an arbitrary constant and k the degree of homogeneity. For an  power-law  with integer exponent $\mathcal{O}(x)=x^{\rm p}$, k=p, and for a linear mapping k=1. Now, based on this concept we introduce the notion of a log-homogeneous operator. An operator (or function) $\mathcal{O}(x)$ is log-homogeneous if

\begin{equation}
\mathcal{O}(x^{\rm p})={\rm l}_{\kappa}\mathcal{O}(x),
\end{equation}

\noindent where ${\rm l}_{\kappa}$ is a constant. We named this type of operator log-homogeneous because the logarithmic functions satisfies its definition since $\log(x^{\rm p})={\rm p}\log(x)$, i.e. ${\rm l}_{\kappa}=p$. Notice however, that the definition of the log-homogeneous operator is more general since ${\rm l}_{\kappa}$ does not necessarily needs to be equal to p.

Now, although generally considered a function, the weighted p-th order transverse structure function can be considered as an operator $\mathcal{S}_{\rm p}^{\rm w}$ that operates on $u({\bf r})$ and $u({\bf r}+\delta {\bf r})$, i.e. $\mathcal{S}_{\rm p}^{\rm w}=\mathcal{S}_{\rm p}^{\rm w}[\delta r;u({\bf r}),u({\bf r}+\delta {\bf r})]$. We propose, as an initial approximation, that $\mathcal{S}_{\rm p}^{\rm w}[\delta r;u({\bf r}),u({\bf r})+\delta {\bf r})]$ is a log-homogeneous operator for the given spatial distribution of the centroid velocity field $U$ of the Huygens Region.\footnote{i.e., the homogeneity of $\mathcal{S}_{\rm p}^{\rm w}$ depends on the homogeneity of the velocities $u\in U$} We make this approximation inspired on the fact that random binary maps where one of the two present values is 0 and the other one is an arbitrary real number distinct from 0, e.gr. 1, capture roughly the contrast in velocity that exist in turbulent velocity fields, in particular, of the field $U$ of the Huygens Region. So, we will assume that log-homogeneity applies, later we will analyse its consequences and later evaluate how much log-homogeneous is the field $U$ of the Huygens Region to later improve our model of the type of homogeneity that $U$ exhibits.

Under the log-homogeneous assumption $\mathcal{C}_{\rm p}^{\rm w}(\delta r)$ transforms into

\begin{equation}
\mathcal{C}_{\rm p}^{\rm w}(\delta r) =\sum_{\rm m,n}^{\rm M'',N''}\sum_{\theta=0}^{2\pi} {\rm w}(\theta;\delta r) \sum_{{\rm i}=1}^{\rm p-1} (-1)^{\rm i} \binom{\rm p}{{\rm i}} l_{\rm p-i,i} u({\bf r}) u({\bf r+}\delta {\bf r}). \label{eqlogCp}  
\end{equation}

As an initial step, we will simplify further assuming that all $l_{\rm p-i,i}=k_{\rm p}/k_2$. This implies that all $C_{\rm p-i,i}^{\rm w}$ are exactly identical to each other and normalised to a maximum value of 1 after normalisation by $k_{\rm p}/2$. This assumption and its consequences holds exactly  for random binary maps, we have proven this analytically and numerically using binarised maps generated from the uniform distribution and the Gaussian distribution. Thus, under the previous assumption we have 

\begin{equation}
\mathcal{C}_{\rm p}^{\rm w}(\delta r) =\frac{k_{\rm p}}{k_2}\sum_{\rm m,n}^{\rm M'',N''}\sum_{\theta=0}^{2\pi} {\rm w}(\theta;\delta r) \sum_{{\rm i}=1}^{\rm p-1} (-1)^{\rm i} \binom{\rm p}{{\rm i}} u({\bf r}) u({\bf r+}\delta {\bf r}). \label{eqlogCpleq}  
\end{equation}

Now, since

\begin{equation}
\sum_{{\rm i}=1}^{\rm p-1} (-1)^{\rm i} \binom{\rm p}{{\rm i}}=-2
\end{equation}

\noindent for arbitrary even integer p, we have\footnote{since $C_{1,1}(\delta r)=C_{\square}^{\rm w}(\delta r)$, i.e. we obtain the linear circular correlation function.} 

\begin{equation}
\mathcal{C}_{\rm p}^{\rm w}(\delta r) =-2 \frac{k_{\rm p}}{k_2}\sum_{\rm m,n}^{\rm M'',N''}\sum_{\theta=0}^{2\pi} {\rm w}(\theta;\delta r)  u({\bf r}) u({\bf r+}\delta {\bf r})=-2\frac{k_{\rm p}}{k_2}C_{\square}^{\rm w}(\delta r). \label{eqlogCpleq1}  
\end{equation}

\noindent Then, Eq.~(\ref{eqkpCp}) reads

\begin{equation}
	S_{\rm p}^{\rm w}(\delta r) =k_{\rm p} - 2\frac{k_{\rm p}}{k_2}\mathcal{C}_{\square}^{\rm w}(\delta r),\label{eqSpbin}
\end{equation}

\begin{figure*}
	\centering
	\includegraphics[scale=0.32]{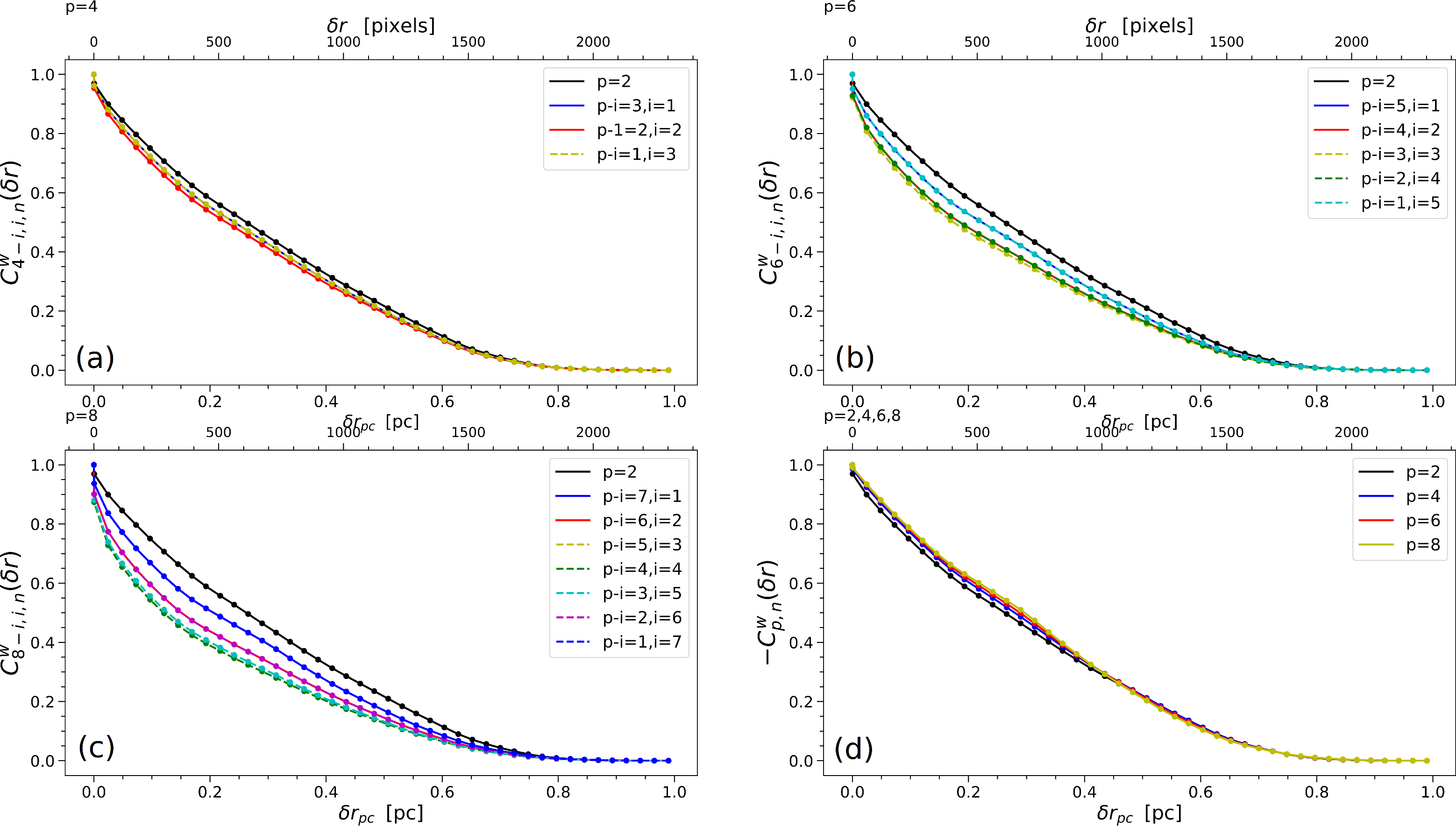}\\
	\caption{Normalised correlations of orders p=2,4,6 and 8. We show the normalised correlations $C_{\rm p-i,i,n}^{\rm w}$ for distinct values of i and also the normalised binomially weighted correlation $C_{\rm p}^{\rm w}/k_{\rm p}$. The black lines correspond to $C_{\rm 1,1,n}^{\rm w}$. Notice the similarity among the  $(C_{\rm p}^{\rm w}/k_{\rm p})$'s, Panel (d). }
	\label{fig:fig11}
\end{figure*}

The last Equation implies that after normalisation by $k_{\rm p}$ all structure functions are identical to $S_{2,\rm n}^{\rm w}(\delta r)$. Hence, This simple idealised model explains roughly the similarity of the even-order structure functions shown in Figures~\ref{fig:fig5}~and~\ref{fig:fig8}. This is the case because the centroid velocity map of the Huygens Regions has almost a log-normal distribution. Our simple model also supports the claim that heavy tails in the velocity PDF --which supply the velocity contrast just as the non-vanishing values in random binary maps-- are an indicator of ongoing turbulence, see \citet{Federrath2013}. However, although this simple initial model captures the essential explanation of the similarity of the even-order structure functions presented in  Figures~\ref{fig:fig5}~and~\ref{fig:fig8}, the model assumes that all $C_{\rm p-i,i}^{\rm w}$ are identical to $C_{1,1}^{\rm w}=C_{\square}^{\rm w}$ after normalisation by $k_{\rm p}/2$ and $k_2/2$, respectively, which is only an approximation for the case of the field $U$ of the Huygens Region. In Figure~\ref{fig:fig11} we present the actual normalised p-th order correlation functions corresponding to the Huygens Region, given by

\begin{equation}
     C_{\rm p-i,i,n}^{\rm w}=\frac{C_{p-i,i}^{\rm w}}{k_{\rm p}/2}.
\end{equation}

\noindent We also present the corresponding normalised binomially weighted correlation function given by

\begin{equation}
\mathcal{C}_{\rm p,n}=\frac{\mathcal{C}_{\rm p}}{k_{\rm p}}.
\end{equation}

Notice that each $C_{\rm p-i,i,n}$ varies between 1 to 0 and can be interpreted as a function that gives the p-th order average correlation value --for the entire field $U$ of the Huygens Region-- between the points over a circumference of radius $\delta r$ and the centre of that circumference, i.e. between points at a distance $\delta r$ and the centre from which the distance is measured. Considering, for instance, the normalised linear circular correlation $C_{1,1,n}=C_{\square}^{\rm w}/(k_2/2)$, we find that on the inertial range, such correlation varies from $\sim 83$\% for $\delta r=70$ pixels (0.0301-pc) to $\sim 7$\% for $\delta r=1500$ pixels (0.6450-pc).  This is an interesting result, since after all, the correlation among the velocities is related to the intricacy of the turbulent patterns that catch so much our attention visually.

\begin{figure*}
	\centering
	\includegraphics[scale=0.32]{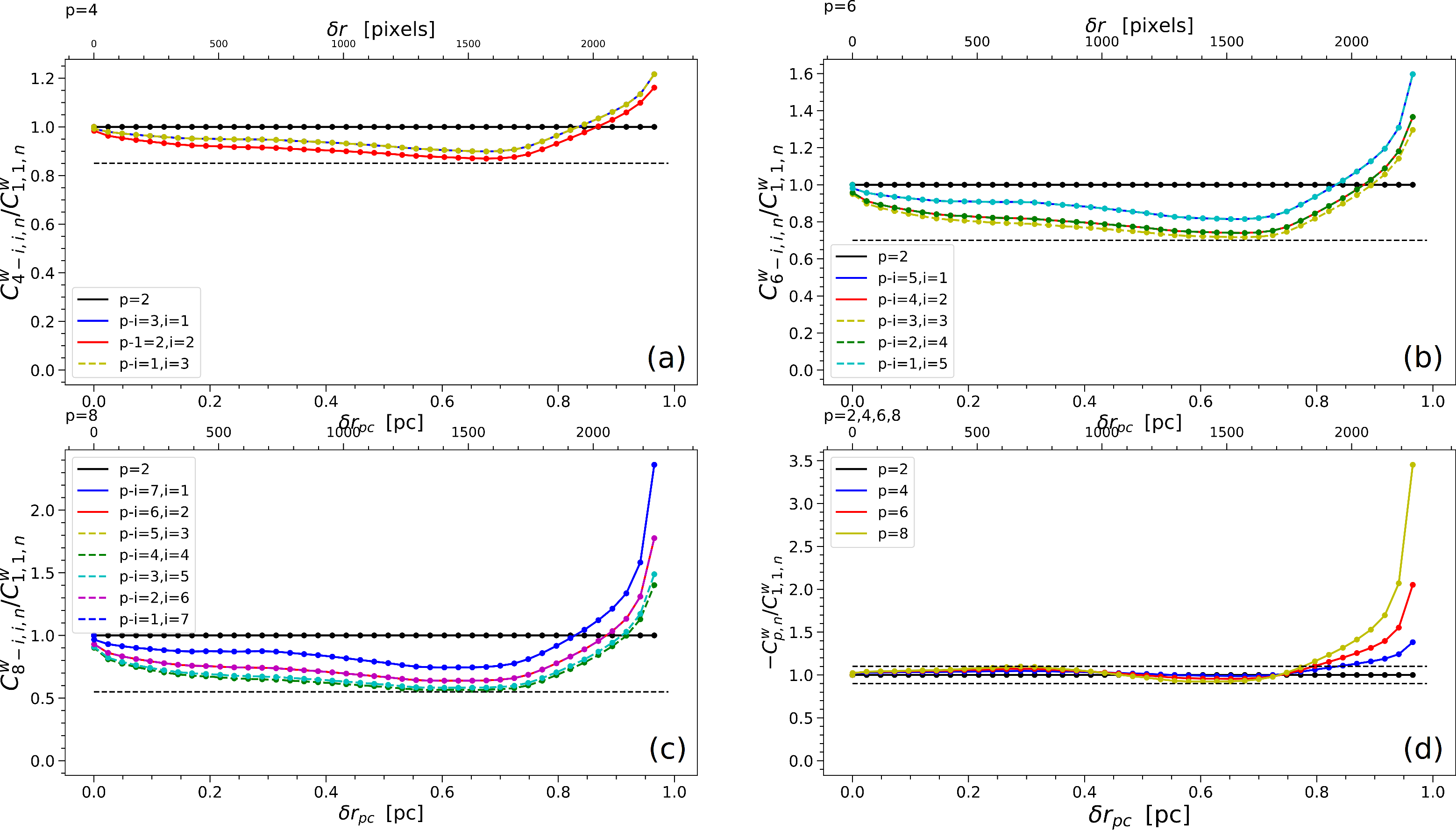}\\
	\caption{Normalised correlations of orders p=2,4,6 and 8 renormalised to $C_{\rm 1,1,n}$. We show the renormalised correlations $C_{\rm p-i,i,n}^{\rm w}$ for distinct values of i and also the renormalised binomially weighted correlation $C_{\rm p,n}^{\rm w}$. Notice the similarity among the  renormalised $C_{\rm p,n}^{\rm w}$'s, Panel (d). The black dashed lines indicate maximum error levels for the $C_{\rm p-i,i}^{\rm w}$ and $\mathcal{C}_{\rm p}^{\rm w}$'s. For the $C_{\rm p-i,i}^{\rm w}$ they are 15\% for p=4, 30\% for p=6, 45\% for p=8 and for the  $C_{\rm p}^{\rm w}$'s they are of  less than 10\%. }
	\label{fig:fig12}
\end{figure*}

In Figure~\ref{fig:fig11}, one can see clearly that the normalised correlation functions $C_{\rm p-i,i,n}^{\rm w}$ are not identical, noticeable small and intermediate deviations from $C_{\rm 1,1,n}^{\rm w}=C_{\square}^{\rm w}/(k_2/2)$ can be observed in the interval corresponding to the inertial range, particularly for p=8, which indicates that the log-homogeneous model with all $l_{\rm p-i,i}$ equal to $k_{\rm p}/k_2$ is only a good approximation but not perfect. On the other hand, one can see that the normalised $\mathcal{C}_{\rm p,n}^{\rm w}$'s are almost identical except for small deviations. This implies that rather than log-homogeneous, the turbulent field $U$ is only quasi-log-homogeneous with a binomially weighted log-homogeneous model describing much better is statistical properties. In order to estimate how much the normalised p-th order correlations deviate from $C_{1,1,n}=C_{\square}^{\rm w}/(k_{\rm p}/2)$, in Figure~\ref{fig:fig12} we present the quotients 

\begin{equation}
	Q_{\rm p-i,i}=\frac{C_{p-i,i,n}^{\rm w}}{C_{1,1,n}^{\rm w}}\label{eqQpi}
\end{equation}

\noindent and also the quotient

\begin{equation}
Q_{\rm p}=\frac{\mathcal{C}_{p,n}^{\rm w}}{C_{1,1,n}^{\rm w}},
\end{equation}

\noindent which will help to estimate the error of the previously presented log-homogeneous model in approximating the true p-th order correlations corresponding to the Huygens Region since the relative errors of the approximations are |1-$Q_{\rm p-i,i}$| and |1-$Q_{\rm p}$| respectively. For p=4,6 and 8 we find maximum relative errors for the $C_{\rm p-i,i,n}$  of 15\%, 30\% and 45\%, respectively. However the relative error in the interval corresponding to the inertial range tends to be less,  between 10\%--30\% in most cases (for  most values of $\delta r$ in the inertial range), see Figure~\ref{fig:fig12}. For the $\mathcal{C}_{\rm p,n}$'s the deviations in the interval corresponding to the inertial range are less than 10\% which indicates the the field homogeneity type is binomially weighted log-homogeneous rather than log-homogeneous, although the log-homogeneous approximation is rather good as we have proven that in most of the cases the deviations from it are  between 10\%--30\%.  The quotients curves seem to blow-up for separation distances $\delta r\sim 2000$ pixels (0.86-pc), but such blow up is irrelevant because it is product of the division of two small quantities (the values of the correlations for such $\delta r$ are less than 0.002 or even smaller, i.e. the value of the correlations are close to zero) and such blow-up occurs when the normalised even order structure functions have values already very close to 1, because as we said, it is only the product of the division of two very small quantities because the values of the correlations are already close to zero.

 We will improve the previous model next. Instead of assuming that all $l_{\rm p-i,i}=k_{\rm p}/k_2$, we will assume that they are the product of slowly varying functions of $\delta r$ with values close to 1 and $k_{\rm p}/k_2$. The exact value of the slowly varying functions for the field $U$ of the Huygens region are given by the quotient functions given by Eq.~(\ref{eqQpi}) and showed in Figure~\ref{fig:fig12}, thus

 \begin{equation}
 \mathcal{C}_{\rm p}^{\rm w}(\delta r) =\sum_{\rm m,n}^{\rm M'',N''}\sum_{\theta=0}^{2\pi} {\rm w}(\theta;\delta r) \sum_{{\rm i}=1}^{\rm p-1} (-1)^{\rm i} \binom{\rm p}{{\rm i}} Q_{\rm p-i,i}\frac{k_{\rm p}}{k_2} u({\bf r}) u({\bf r+}\delta {\bf r}). \label{eqlogCpQ}  
 \end{equation}

\noindent What is interesting, is that the coefficients $Q$ combine in such a manner that

\begin{equation}
	\sum_{{\rm i}=1}^{\rm p-1} (-1)^{\rm i} \binom{\rm p}{{\rm i}} Q_{\rm p-i,i}\approx -2
\end{equation}

\noindent for all even p and almost all $\delta r$ in the inertial range, because of this reason we suggest that the homogeneity type of the field $U$ is a binomially weighted log-homogeneity, since because the former property of the quotients $Q_{\rm p-i,i}$, one recovers almost Eq.~(\ref{eqSpbin}). What does this type of homogeneity means physically? We will answer this question explaining the meaning of a log-homogeneous field interpreting it as a function of space using a 'gain' example after scaling. A log-homogeneous field has nice scaling properties for the correlation functions $C_{\rm p-i,i}^{\rm w}$ and $\mathcal{C}_{\rm p}^w$ and also for the even-order structure functions. These scaling properties just depend on the related  $k_{\rm p}$'s, see Eq.~(\ref{eqkp}). Let's illustrate this with a 'gain' example: if we square the values of the velocities in the field under study, the new linear correlation $C_{1,1}^{\rm w}$'  would only increase by a factor $k_{\rm 4}/k_2$ and the new second-order structure function $S_{2}^{\rm w}$' would increase also by a factor $k_4/k_2$ (where $k_4$ and $k_2$ correspond to the original non-squared field) without changing their normalised spatial profile, i.e. their normalised functional dependency on $\delta r$; if the velocities are elevated to the cubic power then the new linear correlation $C_{1,1}^{\rm w}$' and the new second-order structure function $S_{2}^{w}$' will not change their normalised spatial profile neither, they will only increase by  multiplicative factor of $k_6/k_2$ (where $k_6$ and $k_2$ correspond to the original field not elevated to the third power), and so on. This reproduces the behaviour observed in turbulence experiments and simulations, where a very large increment of the Reynolds number is necessary to observe significant statistical changes on the turbulent field under study. The binomially weighted log-homogeneity is harder to interpret but it has similar scaling properties, in this case, also both the new $\mathcal{C}_{\rm p}^{w}$ and the new $S_{2}^{\rm w};$ scale as $k_{4}/k_2$ or $k_6/k_2$ depending if the field is squared or elevated to the third power. It can be stated  that is a property of the turbulent field $U$ of the Huygens Region that $\mathcal{C}_{\rm p}$, the alternating binomially weighted sum of its associated correlation functions $C_{\rm p-i,i}^{\rm w}$, Eq.~(\ref{eqCp}), is almost identical to $2k_{\rm p}C_{1,1}^{\rm w}/k_2=2k_{\rm p}C_{\square}^{\rm w}/k_2$. Clearly, this type of homogeneity implies certain statistical invariance of the field $U$ when elevated to distinct powers. Up to our knowledge, this is the first time that  quasi-log-homogeneity and binomially weighted log-homogeneity have been presented as properties of a turbulent field.

The general scaling laws for a log-homogeneous field are given below. Let  $\kappa$ be an integer power to which the velocities of the field under study $U$ is elevated. The following scaling laws then apply

\begin{equation}
\begin{array}{ll}
C_{\rm p-i,i}^{\rm w'}=& \frac{k_{p*\kappa}}{k_{\rm p}}C_{\rm p-i,i},\\
S_{\rm p}^{w'}=& \frac{k_{p*\kappa}}{k_{\rm p}} S_{\rm p}^{w}.
\end{array} 
\end{equation}

\noindent For a binomially weighted log-homogeneous field the following scaling laws apply

\begin{equation}
\begin{array}{ll}
\mathcal{C}_{\rm p}^{w'}=& \frac{k_{\rm p*\kappa}}{k_{\rm p}}\mathcal{C}_{\rm p}^{w}, \\
S_{\rm p}^{w'}=& \frac{k_{p*\kappa}}{k_{\rm p}} S_{\rm p}^{w}.
\end{array} 
\end{equation}

\noindent When there are no negative velocities in the turbulent field as in the case of the field $U$ of the Huygens Region, one can set $\kappa$ to any positive real number. When zero velocities are absent too, one can set $\kappa$ to any real number.

\begin{figure}
 	\centering
 	\includegraphics[scale=0.55]{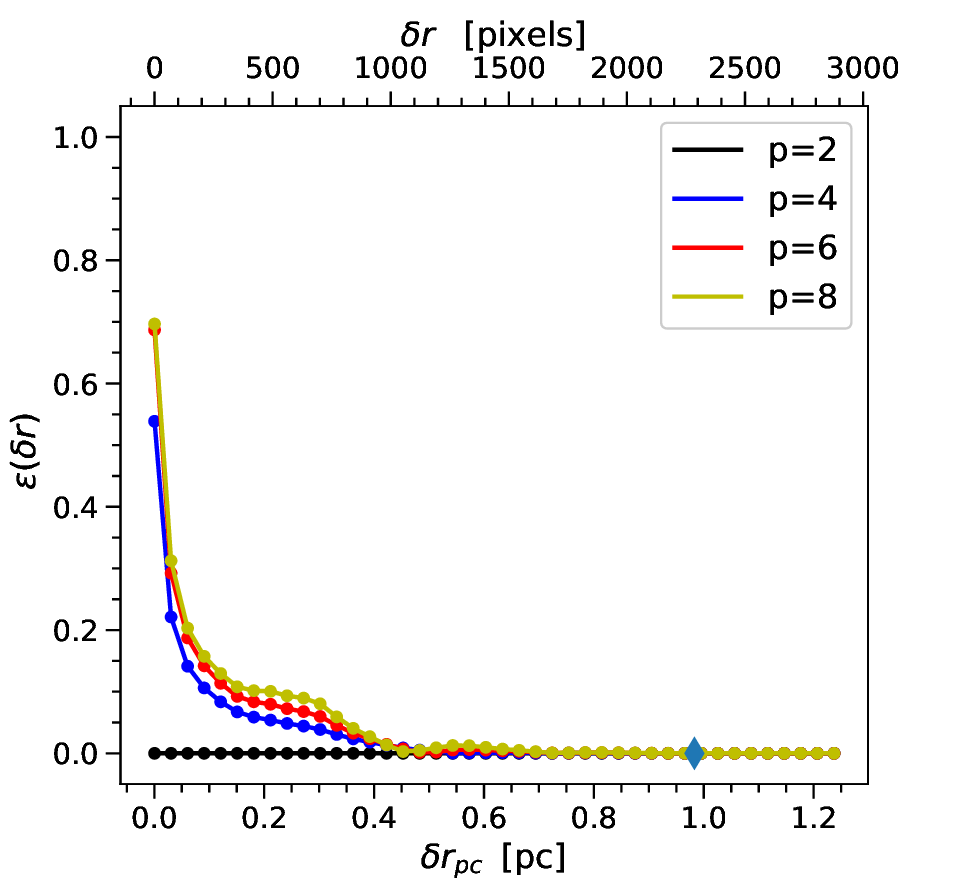}\\
 	\caption{Relative error of the higher order structure functions with respect to $S_{2}^{\rm w}$, $\epsilon(\delta r)=|\frac{S_{2}^{\rm w}-S_{\rm p}^{\rm w}}{S_{2}^{\rm w}}|$. }
 	\label{fig:fig13}
 \end{figure}

The relative errors (deviations from equality) of the higher-order even structure functions can be found from the formula $\epsilon(\delta r)=|(S_{2}^{\rm w}-S_{\rm p}^{\rm w})/S_{2}^{\rm w}|$ and are presented in Figure~\ref{fig:fig13}. In the inertial range, the maximum relative error is $\sim$31\% for p=8 and $\delta r=70$ pixels (0.0301-pc), the relative error decreases from there quickly to reach values of almost zero for $\delta r>800$ pixels (0.344-pc). This demonstrates that the field $U$ is almost binomially weighted log-homogeneous up to order 8th. On the other hand the log-homogeneous approximation has an average relative error of 10\%--30\%. Of course, as p increases the correlations and the structure functions deviate even more from those corresponding to p=2, because of this reason we can talk only of homogeneity up to certain order; in the present communication, we have evaluated the homogeneity up to the 8-th order.

Finally, based on the developments on this Section, it is clear that  the normalised even structure functions can be interpreted as a binomially weighted measure of non-correlation since after normalising Eq.~(\ref{eqSpbin}) by $k_{\rm p}$ it reads

\begin{equation}
S_{\rm p}^{\rm w}(\delta r) =1 - 2\frac{1}{k_2}\mathcal{C}_{\square}^{\rm w}(\delta r)=1-C_{\rm 1,1,n}^{\rm w},\label{eqSpbinnorm}
\end{equation}

\noindent or more precisely, if  we normalize Eq.~(\ref{eqkpCp}) by $k_{\rm p}$ it reads

\begin{equation}
S_{\rm p}^{\rm w}(\delta r) =1+\frac{\mathcal{C}_{\rm p}^{\rm w}(\delta r)}{k_{\rm p}}=1+\mathcal{C}_{\rm p,n},\label{eqkpCpnorm}
\end{equation}

\noindent where $C_{\rm p}$ is negative because of the alternating binomial weighting.\footnote{Notice that the fact that $C_{\rm p}$ is negative does not mean that it measures anti-correlation, it measures positive correlation since each $C_{\rm p-i,i}$ that composes it is positive, $C_{\rm p}$ is negative overall only because of the alternating binomial weighting, so, for the Huygens Region, one must interpret its negative values as a product of an overall negative sign, just as in the case of Eq.~(\ref{eqS2w})} Both the of the above Equations, clearly indicate that $S_{\rm p,n}^{\rm w}$ is a p-th order measure of non-correlation.

\subsection{Similarity of the odd-order structure functions to $S_{\rm 2,n}^{\rm w}(\delta r)$ after normalisation: homogeneity}\label{sec:sec5.2}

The similarity of the odd-order structure functions to $S_{2}^{\rm w}(\delta r)$ cannot be interpreted as exhaustively as in the case of the even-order transverse structure functions because of the absolute value involved in their definition, Eq.~(\ref{eqweighs_gcoords_odd}). However, our numerical calculations show that they have the same behaviour that the even-order transverse structure functions: they differ only from $S_{2}^{\rm w}(\delta r)$ by almost only a constant proportionality factor and they also converge to 1 for $\delta r\ge D$ after normalisation by $k_{\rm p}$.

\subsection{Interpretation of the power-spectrum}\label{sec:sec5.3}

Let's find first a physical explanation for the power-law that corresponds to the initial side of the tail of the power-spectrum. In this case, $\beta=-2.6500$ and it covers the spatial spectral range $n=1.5347$--355 which corresponds to spatial displacements $\delta r\approx 7$--1500 pixels (0.00301--0.6450 pc); however the power-spectrum is not noisy/peaky only in the interval corresponding to the inertial range, $\delta r=70$--1500 pixels (0.0301--0.6450 pc), and thus we consider that the the first power-law is only representative on this latter interval. The following theoretical relationship due to \citet{ZuHone2016} holds between the exponent $\beta_{\rm theory}$ of the initial side of the tail of the power-spectrum and the exponent $\alpha_2$ of the second-order structure function on the inertial range when the power spectrum extracted from the LOS centroid velocity map differs only by a multiplicative constant from the power-spectrum of the associated three-dimensional region, as we are assuming here since the H$\alpha$ lines measured by \emph{MUSE} are optically thin, [see \citet{Miville2003}]

\begin{equation}
	\alpha_2=-(\beta_{\rm theory}+2).
\end{equation}

Thus theoretically, $\beta_{\rm theory}=-2-\alpha_2=-2-0.6868=-2.6868$, which differs slightly from the exponent $\beta=-2.6500$ that we found empirically using regression and our computational geometry based algorithm.  

The exponent $\beta=-0.9747\sim -1$ valid for the end of the tail which corresponds to small spatial scales, $\delta r\sim 1$--5 pixels or $\delta r_{\rm pc}\sim 0.00043$--0.00215 pc, can be interpreted as an indicator of viscous dissipative processes occurring at such small scales, see \citet{Chasnov1998}. Such dissipative processes are probably enhanced by the presence of dust at small scales in the nebula, see \citet{Weilbacher2015}.

\section{Conclusions}\label{sec:sec6}

We have calculated exhaustively the normalised transverse structure functions up to the 8-th order of the sub-field $U$ of the Huygens Region that exhibits turbulence and contains most of the total power of the LOS centroid velocity map (95.16\% of it). We have found that all higher-order structure functions are almost proportional to the second-order structure function, i.e. that they differ from it almost only by a multiplicative factor.  This type of scaling is similar to that of Burgers turbulence. We have interpreted this fact using a simple mathematical model: we have found that the turbulent field $U$ is quasi-log-homogeneous, or to better degree of approximation, a binomially weighted log-homogeneity is possessed by the field $U$ with a deviation error of less than 10\% between the normalised binomially weighted correlation functions $\mathcal{C}_{\rm p,n}^{\rm w}(\delta r)$ and the circular correlation function $C_{1,1,n}=C_{\square}^{\rm w}/(k_2/2)$, see Section~\ref{sec:sec5.1}.

We also found a normalisation factor $k_{\rm p}$ for the p-th order structure functions which makes them adopt the value of 1 after a separation distance $D$ given by Eq.~(\ref{eqEM}). This fact can be used to verify the correctness of the calculated structure functions. The normalisation factor also allows to compare straigthforwardly the structure functions of different orders p-th of the same field to find scaling laws. The normalisation factor also allows to compare straigthforwardly structure functions of the same order for different fields to test the universality or other properties of turbulence.

We have also obtained and analysed the power-spectrum and found that it has a long tail that can be fitted to two power-laws, one with exponent $\beta=-2.6500$ for the initial side of the tail and one with exponent $\beta=-0.9747$ for the end side of the tail. The first power-law covers the inertial range, covering spatial scales of $\delta r=70$--1500 pixels or $\delta r_{\rm pc}= 0.0301$--0.6450 pc. The second power-law corresponds to small scales of $\delta r= 1$--5 pixels or $\delta r_{\rm pc}= 0.00043$--0.00215 pc. We find that the first power-law with exponent $\beta=-2.65$ is consistent theoretically with the assumption that the power-spectrum of the 2D LOS centroid velocity map is proportional to the three-dimensional power-spectrum of the Huygens Region. This is due to the fact that the H$\alpha$ lines measured using \emph{MUSE} by \citet{Weilbacher2015} are optically thin. We also interpret the second power-law with exponent $\beta=-0.9747$ as an indicator of viscous dissipative processes associated to the presence of dust up to the smallest scales in the nebula.

We also presented our real-space weighted algorithm for calculating structure functions and correlation functions. Our algorithm improves previous algorithms as it is based on exact computational geometry derivations. We think that to provide an exact robust algorithm for calculating structure functions and correlation functions is important because after a wide revisitation of the literature, we have found many miscalculated structure functions that do not possess the characteristics that we determined through our exact analysis. 

In resume, to calculate $S_{\rm p}^{\rm w}(\delta r)$ for a field $U$ the next steps must be followed:

\begin{enumerate}
	\item Read the velocity field $U$ and store it in a 2D Cartesian array, if necessary reduce the size I$\times$J of the databox to the tightest databox of size I'$\times$J'  containing $U$, this is recommendable when there are many zeroes surrounding non-vanishing data as the computational time can be reduced significantly
	\item Zero-padd the field $U$ using Eq. (\ref{eqZP1}). For a plot that displays the perfectly flat plateau use $\delta r^{**}>D$. Remember that $D\le L$.
	\item After implemention of Eqs.~(\ref{eqweighs_gcoords})~and~(\ref{eqweighs_gcoords_odd})  calculate $S_{\rm p}^{\rm w}(\delta r')$ for all $\delta r'\le \delta r^{**}$ by varying the integration (summation symbol) limits using the formulae in Eq. (\ref{eqMN}).
\end{enumerate}

For calculating the power spectrum  of a field $U$ proceed as follows:

\begin{enumerate}
	\item Follow step (i) of the previous list
	\item Zero-pad the field $U$ using Eq.~(\ref{eqZP3}) or (\ref{eqZP5}).
	\item Carry out the Fourier transform the resulting zero-padded field and multiply later by its complex conjugated, Eq. (\ref{eqE})
	\item Average circularly with respect to the centre of the resulting field using our weighted scheme that uses exact circular averaging based on computational geometry calculations. These steps will produce the most optimally possible power-spectrum
\end{enumerate}

We have implemented our algorithm for calculating $S_{\rm p}^{\rm w}(\delta r)$ in a code named \texttt{SFIRS} (\emph{Structure Functions calculated in Real Space}) that optimises to the maximum the calculations and that is available under request. \texttt{SFIRS} has also implemented an algorithm for the calculation of the power-spectrum and the correlation functions using exact circular averaging based on computational geometry derivations.

\section*{Acknowledgements}

We thank Dr. P. Weilbacher for given permission of reproducing figure 28(a) from \citet{Weilbacher2015} and also for answering our questions about the data related to the centroid velocity map of the Huygens Region. The results presented in this Paper were calculated using data measured and derived by \citet{Weilbacher2015} and are based on observations collected at the European Southern Observatory under \emph{ESO} programme(s) 60.A-9100(A). We also thank our anonymous referee whose valuable comments helped to significantly improve this Paper.

\section*{Data availability}

The data underlying this article will be shared on reasonable request to the author. The H$\alpha$ LOS centroid velocity map of the Huygens Region which is analysed on this article was obtained by \citet{Weilbacher2015} and is freely available at the web address http://muse-vlt.eu/science/m42/

%%%%%%%%%%%%%%%%%%%%%%%%%%%%%%%%%%%%%%%%%%%%%%%%%%



%\bibliographystyle{mnras}

\begin{thebibliography}{99}
	
\bibitem[\protect\citeauthoryear{Anorve-Zeferino}{2009}]{AnorveZeferino2009}
Anorve-Zeferino G.~A., 2009. MNRAS, 394, 1284
		
\bibitem[\protect\citeauthoryear{Anorve-Zeferino}{2019}]{AnorveZeferino2019}
Anorve-Zeferino G.~A., 2019, MNRAS, 483, 704	

\bibitem[\protect\citeauthoryear{Babiano, Basdevant \& Sadourny}{1985}]{Babiano1985}
Babiano A., Basdevant C. \& Sadourny R., 1985, J. Atmos. Sci., 42, 941	

\bibitem[\protect\citeauthoryear{Batchelor}{1949}]{Batchelor1949}
Batchelor G.~K., 1949, Proc. Royal Soc. London A, 195, 513	

\bibitem[\protect\citeauthoryear{Batchelor}{1951}]{Batchelor1951}
Batchelor G.~K., 1951, Math. Proc. Cam. Phil. Soc., 47, 359

\bibitem[\protect\citeauthoryear{Boldyrev}{2002}]{Boldyrev2002}
Boldyrev S., 2002, ApJ, 569, 841

\bibitem[\protect\citeauthoryear{Boneberg et al.}{2015}]{Boneberg2015}
Boneberg D.~M. et al., 2015, MNRAS, 447, 1341

\bibitem[\protect\citeauthoryear{Brunt \& Mac Low}{2004}]{Brunt2004}
Brunt C.~M. \& Mac Low M.-M, 2004, ApJ, 604, 196

\bibitem[\protect\citeauthoryear{Champeney}{1973}]{Champeney1973}
Champeney D.~C., 1973, {\it Fourier transforms and their physical applications}, pp. 68--76, CUP

\bibitem[\protect\citeauthoryear{Chasnov}{1998}]{Chasnov1998}
Chasnov J.~R., 1998, Phys. Fluids, 10, 1991

\bibitem[\protect\citeauthoryear{Clerc et al.}{2019}]{Clerc2019}
Clerc N. et al., 2019. A\&A, 629, 143	

\bibitem[\protect\citeauthoryear{Cucchetti et al.}{2019}]{Cucchetti2019}
Cucchetti E. et al., 2019, A\&A, 629, 144

\bibitem[\protect\citeauthoryear{Federrath et al.}{2010}]{Federrath2010}
Federrath C. et al., 2010, A\&A, 512, 81	

\bibitem[\protect\citeauthoryear{Federrath}{2013}]{Federrath2013}
Federrath C., 2013, MNRAS, 436, 1245

\bibitem[\protect\citeauthoryear{Frisch}{1995}]{Frisch1995}
Frish U., 1995, MNRAS, {\it Turbulence: the legacy of A.~N. Kolmogorov}, CUP, pp. 142

\bibitem[\protect\citeauthoryear{Kolmogorov}{1991}]{Kolmogorov1991}
Kolmogorov A.~N., 1991, Proc. Royal Soc. A, 434, 9

\bibitem[\protect\citeauthoryear{Konstandin et al.}{2012}]{Konstandin2012}
Konstandin L. et al., JFM, 2012, 692, 183

\bibitem[\protect\citeauthoryear{Miville-Desch\^enes, Levrier \& Falgarone}{2003}]{Miville2003}
Miville-Desch\^enes M.-A., Levrier F. \& Falgarone E., 2003, ApJ, 593, 831

\bibitem[\protect\citeauthoryear{Monin \& Yaglom}{1971}]{Monin1971}
Monin A.~S. \& Yaglom A.~M., 1971, {\it Statistical Fluid Mechanics: Mechanics of Turbulence, Vol. 2},, The MIT Press, pp. 83--85 
	
\bibitem[\protect\citeauthoryear{Padoan et al.}{2003}]{Padoan2003}
Padoan P. et al., 2003, ApJ, 2003, 583, 308

\bibitem[\protect\citeauthoryear{She \& Leveque}{1994}]{She1994}
She Z-S. \& Leveque E., 1994, Phys. Rev. Letters, 72, 336

\bibitem[\protect\citeauthoryear{Schumacher \& Eckhardt}{1994}]{Schumacher1994}
Schumacher J. \& Eckhardt B., 1994, Phys. Plasmas, 1994, 6, 3477

\bibitem[\protect\citeauthoryear{Schmidt, Federrath \& Klessen}{2008}]{Schmidt2008}
Schmidt W., Federrath C. \& Klessen R., 2008, Ph. Rev. Letters., 101, 4505


\bibitem[\protect\citeauthoryear{Schulz-Dubois \& Rehberg}{1981}]{Schulz-Dubois1981}
Schulz-Dubois E.~0. \& Rehberg I., 1981, Appl. Phys., 24, 323

\bibitem[\protect\citeauthoryear{Stewart \& Federrath}{2022}]{Stewart2022}
Stewart M. \& Federrath C., 2002, MNRAS, 509, 5237	

\bibitem[\protect\citeauthoryear{Thomson}{1988}]{Thomson1988}
Thomson D.~J., 1988, Brunel University Thesis, pp. 22--24, London, UK	

\bibitem[\protect\citeauthoryear{Weilbacher et al.}{2015}]{Weilbacher2015}
Weilbacher P.~M. et al., 2015, A\&A, 582, 114 

\bibitem[\protect\citeauthoryear{Xie}{2021}]{Xie2021}
Xie J.-H., 2021, Acta Mech. Sin., 37, 47 


\bibitem[\protect\citeauthoryear{ZuHone, Markevitch \& Zhuravleva}{2016}]{ZuHone2016}
ZuHone J.~A, Markevitch M. \& Zhuravleva I., 2016, ApJ, 817, 110 




\end{thebibliography}


%%%%%%%%%%%%%%%%%%%%%%%%%%%%%%%%%%%%%%%%%%%%%%%%%%

%%%%%%%%%%%%%%%%% APPENDICES %%%%%%%%%%%%%%%%%%%%%

\appendix

\section{Errors due to the lack of necessary and sufficient zero-padding} \label{sec:secA}

\begin{figure}
	\centering
	\includegraphics[scale=0.5]{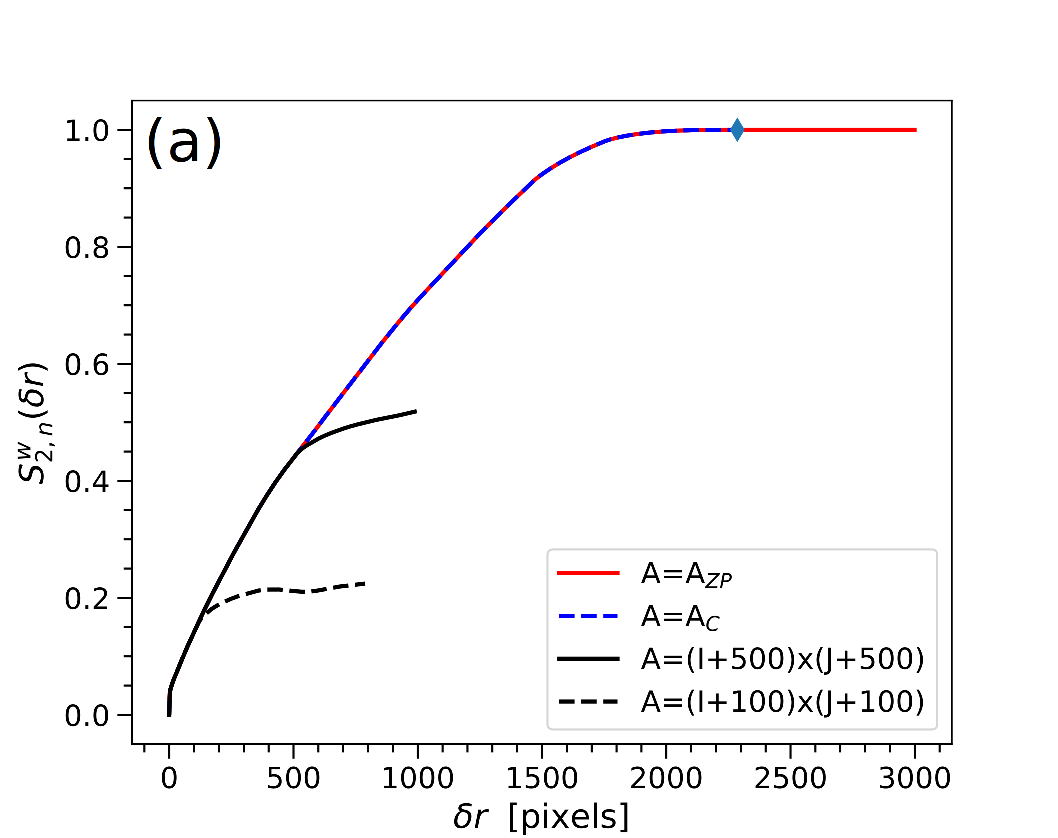}\\
	\includegraphics[scale=0.5]{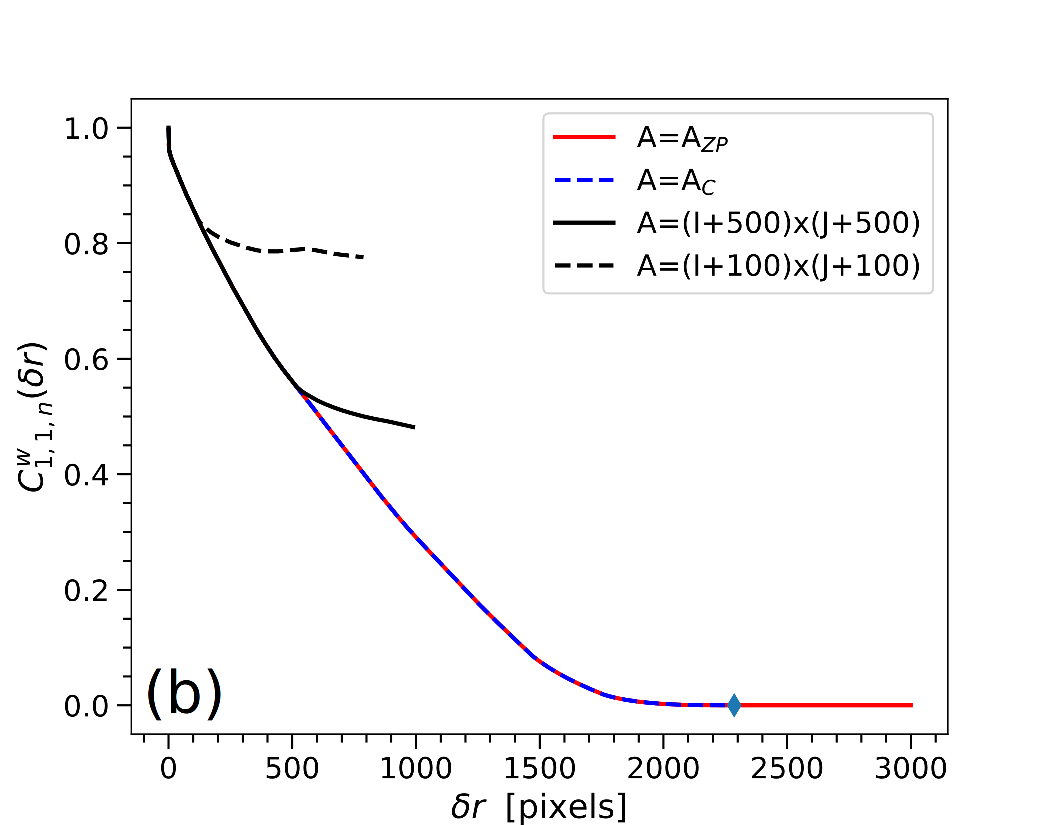}
	\caption{Panel (a): normalised second-order structure functions for the field $U$ of the Huygens Region calculated with different amounts of zero padding. $A$ indicates the area in pixels$^2$ of the extended databox that was used. The blue diamond has the same meaning than in Figure~\ref{fig:fig5}. Panel (b): linear-order circular correlation function calculated with different amounts of zero padding. Clearly, the second-order structure functions and correlations calculated with less zero-padding than indicated by Eq.~(\ref{eqZP3})  are only correct up to $\delta r$ similar to the amount of zero-padding added, after that they present large deviations from the true corresponding functions.}
	\label{fig:figA1}
\end{figure}

We have intentionally calculated the normalised second-order structure function and the normalised circular correlation function through the Fourier-transform-based algorithm using less zero-padding that the minimum optimal zero-padding specified in Eq.~(\ref{eqZP3}). We have used  zero-padding amounts of only 100 pixels and 500 pixels instead of that indicated by Eq.~(\ref{eqZP3}). The results of the calculation using Eq.~(\ref{eqS2w}) are show in Figures~\ref{fig:figA1}(a)-(b). One can see that the second-order structure function and the linear-order circular correlation function are only correct up to $\delta r\sim 100$ pixels when a zero-padding of 100 pixels is used. Similarly, they are correct only up to $\delta r\sim 500$ pixels when a zero-padding of 500 pixels is used. On the other hand, they deviate a lot from the true functions after those spatial scales. This occurs because border effects degrade $S_{2,n}^{\rm w}(\delta r)$ and $C_{1,1,n}^{\rm w}=C_{\square} ^{\rm w}(\delta r)$ for larger spatial scales. The same occurs for real space algorithms when less zero-padding than indicated by Eq.~(\ref{eqZP1}) is used. We have found in the literature many structure functions and correlation functions similar to those of the black lines in Figures~\ref{fig:figA1}(a)-(b). Clearly, they are incorrect but the reasons for this may be difficult to determine but it is possible that a deficit of zero-padding is involved. Notice that the second-order structure functions with less zero-padding than indicated by Eq.~(\ref{eqZP3}) can even display oscillating pseudo-plateaus, we have seen many second-order structure functions reported like this in the literature, so the hypothesis that a lack of proper zero-padding is involved is plausible.

%%%%%%%%%%%%%%%%%%%%%%%%%%%%%%%%%%%%%%%%%%%%%%%%%%

% Don't change these lines
\bsp	% typesetting comment
\label{lastpage}
\end{document}